\documentclass[acmsmall, screen]{acmart}

\AtBeginDocument{%
  }

\usepackage{hyperref}
\usepackage{url}
\usepackage{comment}
\usepackage{wrapfig}
\usepackage{amsmath,mathtools}
\usepackage{soul} 
\usepackage{graphicx}    
\usepackage{caption}     
\usepackage{subcaption}  
\usepackage{enumitem}
\usepackage{booktabs}    
\usepackage{multirow}    
\usepackage{tabularx}    
\usepackage{array}       
\newcolumntype{L}{>{\raggedright\arraybackslash}X} 
\usepackage{xcolor}
\usepackage{adjustbox}
\usepackage{algorithm}
\usepackage{algpseudocode}
\usepackage{xspace}
\usepackage[most]{tcolorbox}
\usepackage{balance}
\usepackage[T1]{fontenc}

\newtcolorbox{findingsbox}{
  colback=gray!5,      
  colframe=black!50,   
  boxrule=0.8pt,       
  arc=3pt,             
  left=8pt, right=8pt, top=6pt, bottom=6pt,
  before skip=10pt, after skip=10pt,
}
\newif\ifcomments
\commentstrue              

\algrenewcommand\algorithmicrequire{\textbf{Input:}}
\algrenewcommand\algorithmicensure{\textbf{Output:}}
\algrenewcommand{\algorithmiccomment}[1]{\hspace{.5em}\(\triangleright\)\,#1}
\algtext*{EndIf}
\algtext*{EndFor}
\algtext*{EndWhile}
\newcommand{\approach}{\textsc{Graphectory}\xspace}
\newcommand{\lang}{\textsc{Langutory}\xspace}
\newcommand{\SA}{SWE-agent\xspace}
\newcommand{\OH}{OpenHands\xspace}

\newcommand{\cmt}[3]{\ifcomments\textcolor{#1}{\textbf{[#2:}~#3\textbf{]}}\fi}
\newcommand{\MRColor}{black}

\newcommand{\shuyang}[1]{\cmt{magenta}{Shuyang}{#1}}

\begin{document}

\acmDOI{10.1145/3798271}
\acmYear{2026}
\acmJournal{PACMPL}
\acmVolume{10}
\acmNumber{OOPSLA1}
\acmArticle{163}
\acmMonth{4}
\received{2025-10-10}
\received[accepted]{2026-02-17}
\setcopyright{cc}
\setcctype{by}

\title{Process-Centric Analysis of Agentic Software Systems}

\author{Shuyang Liu}
\orcid{0009-0009-7264-268X}
\affiliation{%
  \institution{University of Illinois Urbana–Champaign}
  \country{USA}
}
\email{sl225@illinois.edu}
\author{Yang Chen}
\orcid{0009-0008-1409-4163}
\affiliation{%
  \institution{University of Illinois Urbana–Champaign}
  \country{USA}
  }
\email{yangc9@illinois.edu}
\author{Rahul Krishna}
\orcid{0000-0002-5899-6651}
\affiliation{%
  \institution{IBM Research}
  \country{USA}
  }
\email{rkrsn@ibm.com}
\author{Saurabh Sinha}
\orcid{0000-0003-4092-2643}
\affiliation{%
  \institution{IBM Research}
  \country{USA}
  }
\email{sinhas@us.ibm.com}
\author{Jatin Ganhotra}
\orcid{0000-0001-6212-0356}
\affiliation{%
  \institution{IBM Research}
  \country{USA}
  }
\email{jatinganhotra@us.ibm.com}
\author{Reyhaneh Jabbarvand}
\orcid{0000-0002-0668-8526}
\affiliation{%
  \institution{University of Illinois Urbana–Champaign}
  \country{USA}
  }
\email{reyhaneh@illinois.edu}

\begin{abstract}

Agentic systems are modern software systems: they consist of orchestrated modules, expose interfaces, and are deployed in software pipelines. Unlike conventional programs, their execution, i.e., trajectories, is inherently stochastic and adaptive to the problems they are solving. Evaluation of such systems is often outcome-centric, i.e., judging their performance based on success or failure \emph{at the final step}. This narrow focus overlooks detailed insights about such systems, failing to explain how agents reason, plan, act, or change their strategies. Inspired by the structured representation of conventional software systems as graphs, we introduce \approach to systematically encode the temporal and semantic relations in such software systems. \approach facilitates the design of process-centric metrics and analyses to assess the quality of agentic workflows.

Using \approach, we \textcolor{\MRColor}{automatically} analyze $4000$ trajectories of two dominant agentic programming workflows, namely SWE-agent and OpenHands, with a combination of four backbone Large Language Models (LLMs), attempting to resolve SWE-bench Verified issues. Our fully automated analyses \textcolor{\MRColor}{(completed within four minutes)} reveal that: (1) agents using richer prompts or stronger LLMs exhibit more complex \approach, reflecting deeper exploration, broader context gathering, and more thorough validation before patch submission; (2) agents’ problem-solving strategies vary with both problem difficulty and the underlying LLM---for resolved issues, the strategies often follow coherent localization–patching–validation steps, while unresolved ones exhibit chaotic, repetitive, or backtracking behaviors; \textcolor{\MRColor}{and} (3) even when successful, agentic programming systems often display inefficient processes, leading to unnecessarily prolonged trajectories. 

\textcolor{\MRColor}{We also implement a novel technique for real-time construction and analysis of \approach and \lang during the agent’s execution to flag trajectory issues. Upon detecting such issues in the trajectory, the proposed technique notifies the agent with a diagnostic message and, when applicable, rolls back the trajectory. The experimental results show that online monitoring and process-centric analysis, when accompanied by appropriate interventions, can improve resolution rates by $6.9\%$-$23.5\%$ across models for problematic instances, while significantly shortening trajectories with near-zero overhead.} 

\end{abstract}



\maketitle

\section{Introduction}
\label{sec:intro}

Agentic systems powered by large language models (LLMs) have recently emerged as a promising paradigm for autonomously solving complex, multi-step tasks across diverse domains, including software engineering~\citep{shinn2023reflexion,OpenHands,SWE-agent,autocoderoverpaper}, web navigation~\citep{nakano2021webgpt}, scientific discovery~\citep{bran2023chemcrow}, and robotic navigation~\citep{ahn2022doasrobot}. Similar to traditional software systems, agentic systems are built from modules (LLMs, tools, APIs, and memory), and have inputs, internal states, execution logic, and outputs. Instead of a deterministic implementation, however, these software systems operate by producing trajectories---chronological sequences of intermediate reasoning, action, and observation steps---that collectively implement the execution logic for task completion~\citep{yao2022react}. Evaluation of agentic software systems has been mostly outcome-centric: the correctness and quality of the produced trajectory are determined by whether it can solve a problem. 

\begin{figure}[t]
    \centering
    \includegraphics[width=0.98\linewidth]{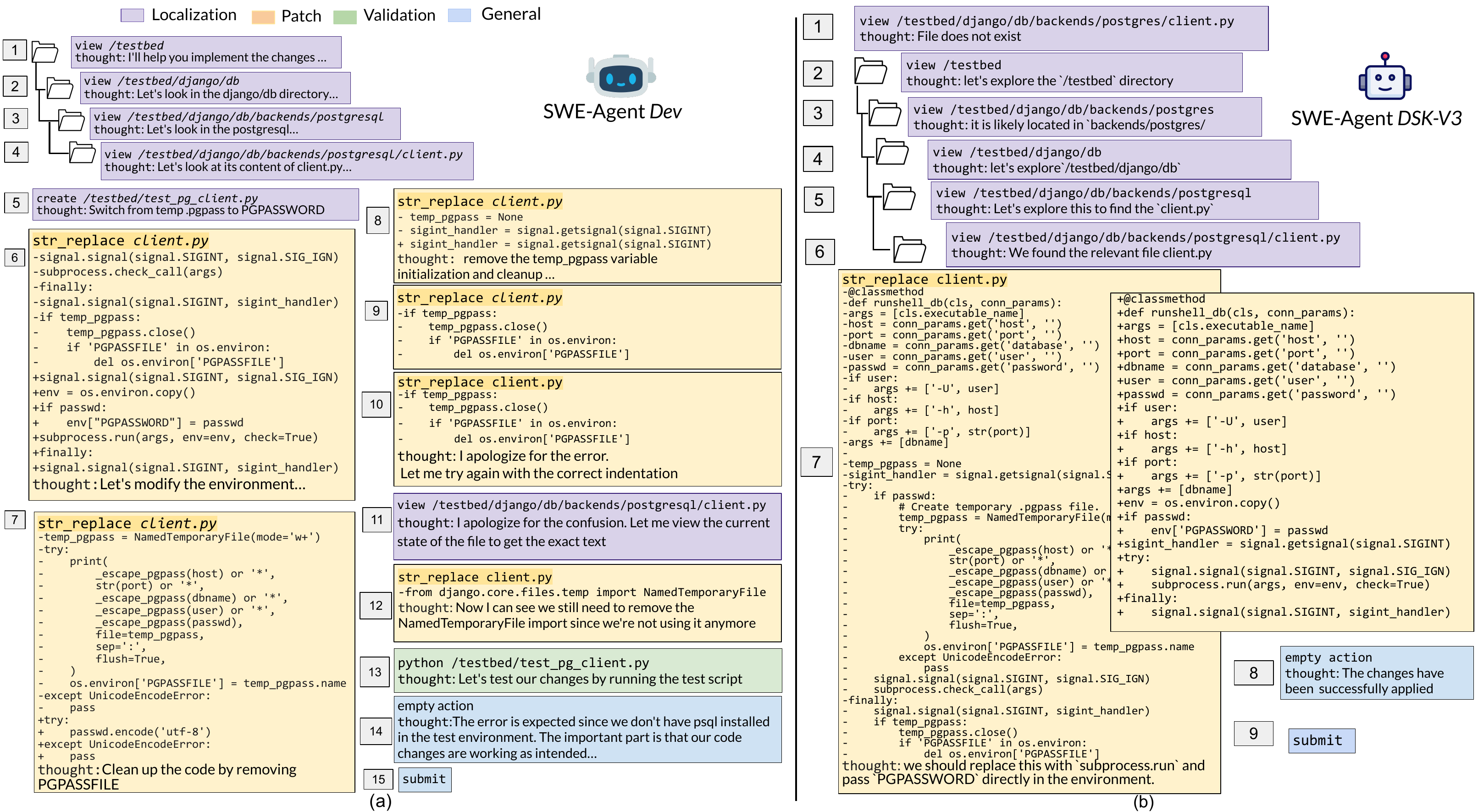}
    \vspace{-8pt}
    \caption{Raw trajectories of (a) \SA\textsubscript{Dev} and (b) \SA\textsubscript{DSK-V3} when resolving  \texttt{\small{django-10973}}}
    \vspace{-5pt}
    \label{fig:illustrative-example}
\end{figure}

An outcome-centric evaluation overlooks the intermediate steps that lead there, masking recurrent inefficiencies and preventing us from understanding whether success came from systematic reasoning or by chance. 
For example, consider the trajectory traces of \SA~\citep{SWE-agent} with Devstral-Small (Figure~\ref{fig:illustrative-example}-a) and DeepSeek-V3 (Figure~\ref{fig:illustrative-example}-b) for repairing the issue \texttt{\small{django-10973}} in SWE-bench~\cite{jimenez2023swe}. From an outcome-centric perspective, both agents, \emph{\SA\textsubscript{Dev}} and \emph{\SA\textsubscript{DSK-V3}}, successfully repair this issue, and the patches are almost similar. However, their trajectories, i.e., the sequence of reasoning about how to solve the problem and taking the appropriate actions, are different: 
    \SA\textsubscript{Dev} starts by localizing the bug to \texttt{\small{client.py}} file (steps 1--4), creating a reproduction test (step 5), and editing multiple locations of the \texttt{\small{client.py}} (steps 6--8). After two additional repetitive failed edits (step 9--10), it re-views and edits a different block of \texttt{\small{client.py}} file (steps 11 and 12), executes the previously created reproduction test (step 13) to validate the patch, and concludes with patch submission (steps 14-15). \SA\textsubscript{DSK-V3}, also starts the process by localizing the bug to \texttt{\small{client.py}} file (steps 1--6), but it only edits the file once (step 7) and prepares the final submission of the patch without any validation (steps 8--9).
Despite final success, both agents suffer from several strategic pitfalls: 

\begin{enumerate}[leftmargin=*]
    \item \SA\textsubscript{Dev} edits \texttt{\small{client.py}} line by line in steps 6--8, followed by repeating a failed edit twice in steps 9--10, and the final edit at step 12, compared to \SA\textsubscript{DSK-V3}, which generates the patch in one attempt at step 7. 
    \item \SA\textsubscript{DSK-V3} explores the project structure instinctively, zooming in (step 1), zooming out (step 2), and zooming in again (steps 3--6), until it localizes the bug, compared to \SA\textsubscript{Dev}, which zooms in on the project hierarchy step by step to localize the bug. 
    \item None of the agents run regression tests for patch validation, with \SA\textsubscript{Dev} only checking if reproduction tests pass on the patch (step 13), and \SA\textsubscript{DSK-V3} only considering the correctness of the patch through edge cases (step 8). 
\end{enumerate}

As demonstrated, process-centric analysis, which focuses on trajectories, can provide additional insights about \emph{how} agents solve or cannot solve a given problem. Recently, there have been several attempts at analyzing agentic workflows~\cite{cemri2025multi,deshpande2025trail,liu2025empirical,ceka2025understanding,kartik2025agentcompass}, with a focus on deriving a taxonomy of failure modes through manual, expert analysis of agentic trajectories. While the failure mode taxonomies are valuable, manual analysis is subject to human bias and incompleteness, and also does not scale to analyses of new agentic systems. More importantly, trajectories in their raw format only represent steps as a linear sequence in chronological order, failing to effectively capture \emph{semantics} of agentic behavior, such as execution flow and problem-solving strategies, and whether they are efficient or follow the expected planning. Systematic analysis of linear trajectories at scale across different problems to determine \emph{common strategies or pitfalls} is also non-trivial.

To promote systematic and scalable process-centric analysis of agent trajectories, this paper proposes \approach, a rich graph that can be automatically generated from linear trajectory logs. The nodes in \approach are agent actions, and two nodes are connected through an edge if (1) one action temporally follows another in the trajectory log or (2) the two actions operate on subsuming entities within the problem space (\S \ref{sec:design}). In addition to \approach, this paper also introduces \lang, a human-readable abstract of \approach that represents \emph{language} of trajectories. \approach enables a systematic characterization and quantitative comparison of trajectories given graph properties (\S \ref{subsec:metrics}). More importantly, \approach and \lang allow a range of process-centric analyses that are not possible with linear trajectory logs, i.e., \emph{Phase Flow Analysis} (\S \ref{subsub:phase-flow-anaysis}) and \emph{Pattern Detection} (\S \ref{subsub:pattern-detection}), supporting systematic evaluation of how agents solve problems and their inefficiencies, not just whether they succeed or fail. 

We study agentic programming systems, namely SWE-agent~\citep{SWE-agent} and OpenHands~\citep{OpenHands}, due to their rich and diverse trajectories: they involve cycles of planning, bug localization, code modification, and patch validation. For our experiments, we run these agentic systems with four LLMs as the backbone (DeepSeek-V3~\citep{deepseek-v3}, DeepSeek-R1~\citep{deepseek-r1}, Devstral-small-2505~\citep{devstral-small}, and Claude Sonnet 4~\citep{claude-sonnet-4}), solving real-world GitHub issues across $12$ repositories from the SWE-Bench Verified benchmark~\citep{swebench-verified}. Our comprehensive systematic analysis reveals that: 

\begin{itemize}[leftmargin=*]
    \item \textbf{Unsuccessful runs are 
    full of inefficiencies.} Across agents and backbone models, \approach of unresolved issues is consistently larger than that of resolved ones, with more back edges, demonstrating more repetitions (\S \ref{subsub:repair-status-analysis}) and inefficient patterns (\S \ref{sec:rq3}). 

    \item 
    \textbf{Trajectory complexity grows with task difficulty.} As problems become harder for human developers, agents also explore deeper and wider (\S \ref{subsub:human-analysis}), with more frequent strategy shifts to sustain progress (\S \ref{subsub:eval-shared-strategy}).

    \item
    \textbf{Adaptive strategy refinement.} Though agents primarily follow the global plan outlined in the system prompt, they also change their strategy at intermediate turns (\S \ref{sec:rq2}). This often requires additional searching and \textcolor{\MRColor}{iterative} debugging for \textcolor{\MRColor}{more complex} problems.

    \item \textbf{Imbalance between efficiency and Success.} Even when successful in issue repair, agents still exhibit significant inefficiency during problem solving (\S \ref{sec:rq3}), showing the importance of process-centric evaluation for highlighting the imbalance between efficiency and repair success.
\end{itemize}

\textcolor{\MRColor}{
\approach and \lang can also be constructed in real time as agents progress along their trajectories to detect inefficiencies and other trajectory issues.
When detected and addressed promptly, this prevents them from affecting the agent’s outcome. To demonstrate the online analysis capabilities of \approach and \lang, we construct/monitor/analyze them during agent execution. Upon detecting an issue in the current trajectory through process-centric analysis, the monitor communicates with the agent, suggesting changes to the current strategy to resolve the issue.
Re-executing the SWE-agent under online monitoring and intervention setting on the instances with problematic trajectories
shows a significant improvement on both trajectory quality and resolution rate across models: out of $86$ instances that \emph{consistently} remained problematic through multiple executions,
online monitoring and intervention execution resulted in non-problematic trajectories in $94.1\%$.
This improvement was achieved with zero cost and minimal time overhead (less than $10$ milliseconds), while improving resolution rate by $11.9\%$, on average (\S \ref{sec:mr-monitor}).}

Our contributions are: (1) a novel structural representation of agent trajectories, i.e., \approach; (2) a novel human-readable abstraction of \approach, representing the language of trajectories, \lang; (3) a series of process-centric metrics that quantify the complexity of trajectories; (4) a series of process-centric analyses to investigate problem-solving strategies and inefficiency patterns; (5) a systematic evaluation of $4000$ trajectories, obtained from \emph{eight} $\langle \text{agent}, \text{model} \rangle$ pairs, providing the community with a rich dataset to explore future process-centric analyses using \approach and \lang; and (5) \textcolor{\MRColor}{a novel approach for online monitoring and intervention of agentic programming trajectories}.

\section{Process-Centric Data Structures, Metrics, and Analyses} 
\label{sec:design}

We formally define \approach below, and will use the notions to define process-centric evaluation metrics (\S \ref{subsec:metrics}) and analysis (\S \ref{subsec:analyses}). To illustrate the definitions and concepts, we use \approach examples obtained from programming agents' trajectories. 

\vspace{3pt}
\noindent \textbf{Definition 1 (\approach).} 
\label{def:approach}
\approach is a cyclic directed graph 
$G = (V, TE \uplus SE)$. 

\begin{itemize}[leftmargin=*]
\renewcommand\labelitemi{-}

  \item Each node $v_x=(k_x,p_x,l_x,S_x,O_x,B_x) \in V$ corresponds to a distinct action taken by the agent in the trajectory to accomplish a task.\footnote{We only consider actions as nodes, and not reasoning, since agents follow the ReAct principle~\cite{yao2022react}, and all actions are due to a prior reasoning. Because reasonings are hard to distinguish, adding generic reasoning nodes is not useful and increases the graph size and complexity.} $k_x$ represents the \emph{unique key} that distinguishes $v_x$ from other nodes as a combination of \emph{action type} and \emph{action arguments}; $p_x$ denotes the problem-specific \emph{logical phase} that the action belongs to; $l_x$ represents the structural navigation level 
  at which the action is taken; $S_x$ is a list of incoming edges to $v_x$, each representing a \emph{trajectory step} $s_{x}^{i}$, at which the action is taken; $O_x$ is a list such that the binary value $o_{x}^{i}$ represents the \emph{outcome of action} at trajectory step $s_{x}^{i}$; and $B_x$ is a list such that the  $b_{x}^{i}$ determines agent's \emph{observations} at trajectory step $s_{x}^{i}$.

  \item A \emph{temporal edge} $e_{x,y}=(v_x,u_y,s_{x}^{i}) \in TE$ encodes a transition at trajectory step $s_{x}^{i}$ from action $v_x$ to action $u_y$ in the chronological execution order. A \emph{structural edge} $e'_{x,y}=(v_x,u_y) \in SE$, where $l_x \ge l_y$, represents a subsuming navigation relationship, which can vary depending on the problem domain: directory $\mapsto$ file $\mapsto$ block in software engineering, room $\mapsto$ zone $\mapsto$ object in robotics, or theory $\mapsto$ hypothesis $\mapsto$ experiment in scientific discovery. $TE$ captures the temporal problem-solving progress, whereas $SE$ demonstrates navigation in the problem space. 
  
\end{itemize}

While \approach provides a rich representation of an agentic software system in-depth semantics analysis, graphs alone can be overly detailed and challenging to compare across agents, backbone LLMs, or problems. To complement the structural view, we define \lang, a compact, human-readable abstraction of the \approach that represents \emph{language} of trajectories.

\vspace{3pt}
\noindent \textbf{Definition 2 (\lang).}
\label{def:lang}
Given a \approach $G = (V, TE \uplus SE)$, the corresponding \lang is defined as \textcolor{\MRColor}{$\mathcal{L}((V, TE \uplus SE), \Phi) = \prod_{v_i \in V} \pi(v_i)$},
where $\pi: V \to \Phi$ is a projection that maps each node $v_i$ to a symbol in the alphabet $\Phi$.

By varying the alphabet $\Phi$ and using different vocabularies, we can examine agents from various perspectives. For example, when the alphabet symbols correspond to logical, problem-specific phases, \lang provides a compact summary of the phase sequences that agents follow while solving problems, and an overview of their problem-solving strategy. This enables \emph{rapid} identification of shared or divergent strategies, and systematic comparison of how different agents structure their problem-solving processes. 

Another important benefit of \lang is the ability to determine agents' planning deviations from the \emph{expected plan} introduced to them by the system's prompts. For example, \SA instructs agents to resolve the issue~\cite{SWE-agent-plan} by finding the code relevant to the issue description, generating a reproduction test to confirm the error, editing the source code, running tests to validate the fix, and thinking about edge cases to ensure the issue is resolved. \OH provides more detailed instructions~\cite{open-hands-plan}: read the problem and reword it in clearer terms, install and run the tests, find the files that are related to the problem, create a script to reproduce and verify the issue before implementing fix, state clearly the problem and how to fix it, edit the source code, test the changes, and perform a final analysis and check. These instructions can serve as the \emph{problem grammar}, providing us a ground truth to check \lang of agents against to determine whether their \emph{local} step-by-step ReAct is correct or not. 

\subsection{\approach and \lang Construction}
\label{graphectory-and-langutory-construction}

\begin{algorithm}[t]
\caption{Phase Labeling of \approach}
\label{alg:phase-mapping}
\small

\algrenewcommand\algorithmiccomment[1]{\hfill$\triangleright$~#1}
\begin{algorithmic}[1]
\Require Phase-agnostic \approach $G' = ((k_x,\bot,S_x,O_x,B_x) \in V, TE \uplus SE))$, Phase Map $map$
\Ensure \approach $G=((k_x,p_x,S_x,O_x,B_x) \in V,, TE \uplus SE))$

\ForAll{$v_x \in G'.V$}

    \If{$|map[k_x.action] == 1|$} \Comment{Checks if $map$ contains a unique phase for the action}
        \State $p_x \gets$ getPhase ($map,k_x.action$)
        \State $continue$
    \EndIf{}

    \State $\text{isTest} \gets$ isTestFile ($v_x.k_x.args$) \Comment{Checks if the action is performed on a test file}

    \State $\text{afterPatch} \gets$ patchingPerformed ($\langle v_0, \ldots, v_{x-1} \rangle$)\Comment{Checks if any patching is done by prior actions}

    \If{$a \in \{\texttt{create},\texttt{str\_replace},\texttt{insert},\texttt{sed}, \texttt{touch}\}$ or redirection}
        \State $p_x \gets$ $\text{isTest} \;?\; (\text{afterPatch} \;?\; V : L) : P$

        \State $continue$
    \EndIf

    \If{$\text{cmd} \in \{\texttt{grep}, \texttt{cat}, \texttt{find}, \texttt{ls}\}$}
        \State $p_x \gets$  $(\text{isTest} \land \text{afterPatch}) \;?\; V : L$ 
        \State $continue$
    \EndIf

    \If{$\text{cmd} \in \{\texttt{python}, \texttt{pytest}\}$}
        \State $p_x \gets$  $\text{afterPatch} \;?\; V : L$ 
        \State $continue$
    \EndIf

    \State $p_x \gets$ general

\EndFor

\State $G \gets G'$

\State \Return $G$

\end{algorithmic}

\end{algorithm}
Given a trajectory log, \approach defines the nodes as trajectory actions. Agents may bundle multiple low-level actions into one composite action~\citep{inaba2023multitoolcot,lu2023chameleon,suris2023vipergpt}. To better capture trajectory semantics, e.g., strategy changes or repeated actions, \approach disaggregates composite actions into their constituent parts. Next, \approach adjusts the edges based on whether an action temporally follows another in the trajectory or the two actions operate on subsuming entities within the problem space. It allows multiple temporal edges between the same node pairs to represent repeated actions at different steps. To reflect the agent's logical workflow, \approach groups nodes into color-coded subgraphs based on logical problem-solving phases, assuming a provided action-phase mapping. If an action does not fall into any phases, it is a 
\emph{general} action.

For programming agents and in the context of the issue repair problem, we follow the best practices in software engineering to determine three logical phases~\cite{long2016automatic,weimer2009automatically,kim2013automatic,liu2020efficiency}: \colorbox[HTML]{D4CCE6}{Localization} phase, which should pinpoint the location of the issue in the code; \colorbox[HTML]{FFE18E}{Patching} phase, which modifies the code at the specified locations to resolve the issue; and \colorbox[HTML]{D4E7CD}{Validation} phase, which ensures the modification resolves the issue without any regression. For the actions, where the corresponding tool or command does not fit into any of these problem-related categories, e.g., \texttt{\small{submit}} command in \SA prepares the patch for submission, we label the phase as \colorbox[HTML]{CFE0F6}{General}.

To map the logical phases to \approach nodes, we adopted the following systematic annotation procedure: two authors and a frontier LLM, i.e., GPT-5,\footnote{We chose GPT-5 as the annotator as it was not used as a backbone LLM in our experiments.} independently labeled each action with at least one of the predefined phases (or \emph{general} if inapplicable). Human annotators relied on domain knowledge and existing tool documentation~\citep{SWE-agent-tools, OpenHands-tools} or command manual (e.g., \texttt{\small{man [action-name]}}) rather than LLMs, to avoid contamination of decisions across annotators. This procedure balances human expertise, automated support, and reproducibility, while mitigating bias from any single annotator. When all annotators agreed, we assigned the consensus label to a given action; otherwise, the two human annotators met to reach consensus, considering the LLM thought process in the discussion.\footnote{The details of labeling by annotators are available in our artifacts.}
All annotators noted that some tools can be used in different phases: 

\begin{itemize}[leftmargin=*]
\renewcommand\labelitemi{-}
    
    \item File editing tools, for example, \texttt{\small{create}}, \texttt{\small{str\_replace}}, and \texttt{\small{insert}} can be used to create or modify a \emph{test} file for the purpose of \emph{reproducing the issue} (Localization), an \emph{application} file for the purpose of fixing the bug (Patching), or a \emph{test} file for the purpose of validating a generated patch (Validation).

    \item Some commands that are primarily used for viewing file contents can also be used to modify or create a file. For example, \texttt{\small{cat [FILE\_PATH]}} tells the shell to read a file identified by the \texttt{\small{FILE\_PATH}}. In its commonly used form, the agents may use it to read an application code to pinpoint the buggy line (Localization phase) 
    or determine the set of proper regression tests to be executed (Validation phase). However, commands such as \texttt{\small{cat}}, \texttt{\small{grep}}, and \texttt{\small{echo}} can use redirection (\texttt{\small{>}} or \texttt{\small{>\,>}}) to create or modify existing files. Depending on the context for creating or modifying a file, the command can be used in Localization, Patching, and Validation, similar to the first scenario. 
    
    \item For debugging, testing 
    can be used for bug localization or validating the generated patch. The agents use \texttt{\small{python}} for executing newly generated reproduction or regression tests, or \texttt{\small{pytest}} for executing existing regression tests. Again, depending on the context, the same tool/command usage can be used for different purposes and, consequently, assigned to different phases. 
        
\end{itemize}

\begin{algorithm}[t]
\caption{\lang Construction}
\label{alg:lang-extraction}
\small
\algrenewcommand\algorithmiccomment[1]{\hfill$\triangleright$~#1}
\begin{algorithmic}[1]
\Require \approach $G = (V, TE \uplus SE)$, alphabet $\Phi = \{L,P,V\}$
\Ensure \lang $\mathcal{L}$


    \State $PO, RL \gets []$, $vocabIndex \gets 1$
    \State $FG \gets$ flattenGraphectory($V$:$V.S$,$TE$) \Comment{Sort nodes according to step index $s_{x}^{i}$ along temporal edges}

    
    \ForAll{$v_x \in FG$}
        \If{$p_x \in \Phi$}
            \State $vocab \gets$ assignVocabulary($FG[x].p_x$,$\Phi$)
            \If{$vocabIndex = 1$ \textbf{or} $vocab \neq PO[vocabIndex-1]$} \Comment{Start or new phase}
                \State $PO[vocabIndex] \gets vocab$
                \State $RL[vocabIndex] \gets 1$
                \State $vocabIndex \gets vocabIndex+1$
            \Else 
                \State $RL[vocabIndex-1] \gets RL[vocabIndex-1]+1$ \Comment{Extend run length}
            \EndIf
        \EndIf
    \EndFor
    
  \State $\mathcal{L} \gets$ constructLanguage($PO$,$RL$)
  \State \Return $\mathcal{L}$

\end{algorithmic}
\end{algorithm}
\begin{figure}[t]
    \centering
    \includegraphics[width=\linewidth]{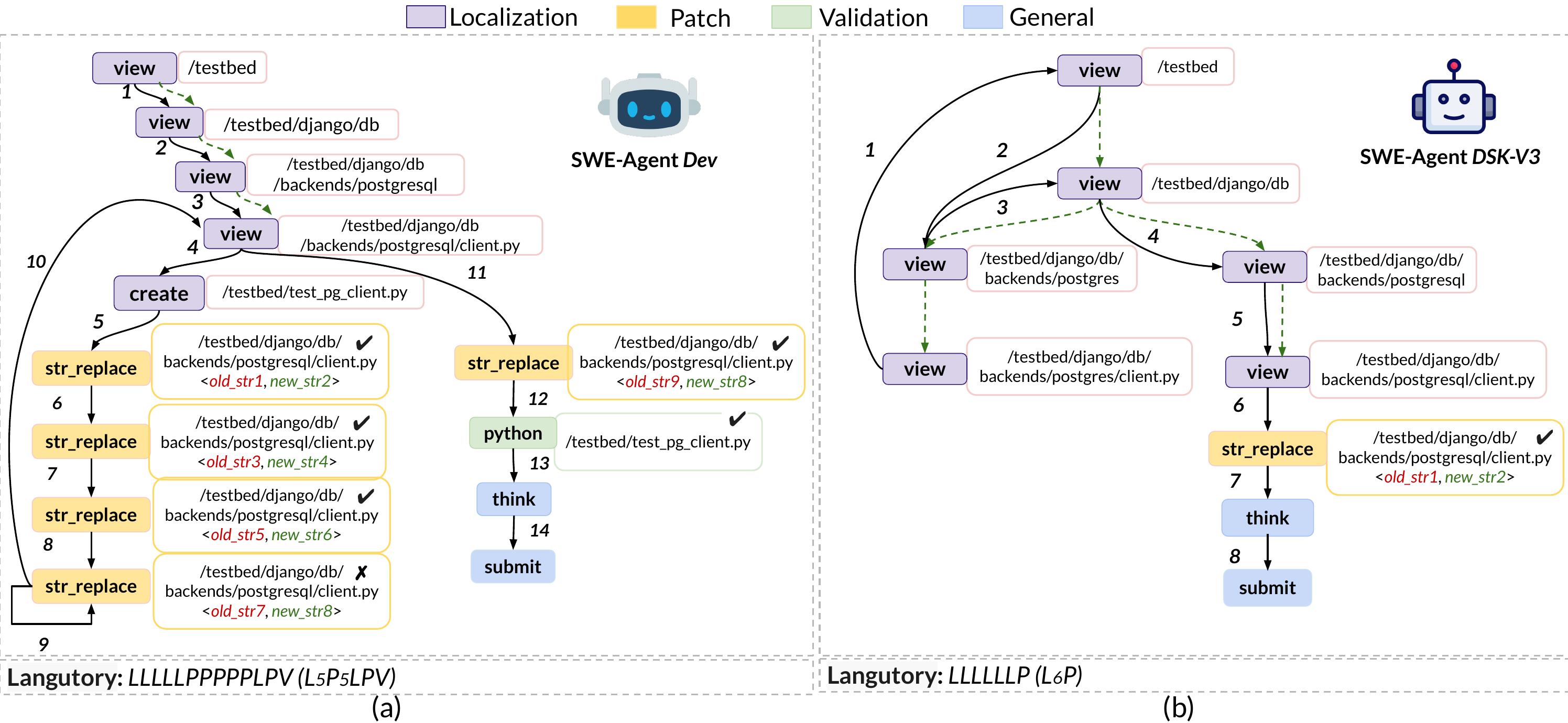}
    \vspace{-15pt}
    \caption{\approach of \SA\textsubscript{Dev} (a) and \SA\textsubscript{DSK-V3} (b) for problem django-10973}
    \vspace{-5pt}
    \label{fig:graphectory-example}
\end{figure}

As a result of the post-annotation discussions, we devise Algorithm~\ref{alg:phase-mapping} for \emph{phase labeling}: given the phase-agnostic \approach $G'$ (with empty phase labels) and the phase map $map$ by annotators, the algorithm goes over all the \approach nodes $v_x$ and directly assigns the phase from $map$, if a unique phase exists for an action (lines $2$--$4$); otherwise, it analyzes 
whether the action is performed on a test file (line $5$) or any patching is done by prior actions (line $6$), and assigns the phase according to the consensus among the annotators (lines $7$--$16$). This algorithm enables the construction of \approach for any agentic programming system. For agents with different tools, the users can update the phase map with minimal overhead compared to the initial effort. 

Algorithm ~\ref{alg:lang-extraction} describes the \lang construction, 
which traverses each node in temporal order and compresses consecutive identical phases using run-length encoding. For example, Figure~\ref{fig:graphectory-example}-a is compressed to $L_5P_5LPV$, with [L,P,L,P,V] as the phase skeleton $PO$, [5,5,1,1,1] as the run length $RL$. Figure~\ref{fig:graphectory-example}-b is compressed to $L_6P$, with [L,P] as the phase skeleton $PO$, [6,1] as the run length $RL$. 
Figure~\ref{fig:graphectory-example} demonstrates the \approach and \lang of \SA\textsubscript{Dev} (Figure~\ref{fig:graphectory-example}-a) and \SA\textsubscript{DSK-V3} (Figure~\ref{fig:graphectory-example}-b) to solve problem \texttt{\small{django-10973}} of SWE-bench. The phases are color-coded as \colorbox[HTML]{D4CCE6}{Localization}, \colorbox[HTML]{FFE18E}{Patching}, \colorbox[HTML]{D4E7CD}{Validation}, and \colorbox[HTML]{CFE0F6}{General}. The solid black lines and green dashed lines demonstrate temporal edges ($TE$) and structural edges ($SE$), respectively. The alphabet of \lang is $\Phi=\{L, P, V\}$, representing the initial of unique logical phases in program repair: \underline{\textbf{L}}ocalization, \underline{\textbf{P}}atching, and \underline{\textbf{V}}alidation. \approach and \lang promptly provide the following insights about the agents' behavior:

\begin{itemize}[leftmargin=*]
\renewcommand\labelitemi{-}
    \item In the \approach of \SA\textsubscript{Dev}, there is a \emph{self-back edge} 9 and \emph{back edge} 10, both indicating potential inefficiencies in agent reasoning or tool usage, resulting in repeating failed edits or refining the localization plan (first pitfall in \S \ref{sec:intro}). 

    \item In the \approach of \SA\textsubscript{DSK-V3}, we observe opposite directions of \emph{structural} edges and \emph{temporal} edges during the bug localization. This immediately suggests inefficiency in the analysis of issue description, causing the agent to go deep and come back and go deep again into the code base hierarchy (second pitfall in \S \ref{sec:intro}). 

    \item Comparing \lang of \SA\textsubscript{Dev} ($L_5P_5LPV$) and \SA\textsubscript{DSK-V3} ($L_6P$) with the expected planning introduced to agents by \SA~\cite{SWE-agent} ($LPV$), we can observe that \SA\textsubscript{DSK-V3} performs no validation on the generated patch, and only \emph{thinks} that the patch resolved the issue. Relying solely on regression tests without performing reproduction testing (or even worse, not performing any testing) is flawed because it verifies only that no new issues are introduced, but fails to confirm whether the original failure has actually been resolved~\cite{le2011genprog,long2016automatic}.
\end{itemize}

Agentic systems, in general, and agentic programming systems, in particular, can be viewed as modern software systems. For classic software systems, the inter-procedural control flow graph (CFG) captures the execution semantics. Similarly, \approach serves as a structured representation of agentic trajectories, where execution of each step forms the functionality of the agents. Inspired by this relationship, we present a series of process-centric metrics (\S \ref{subsec:metrics}) and process-centric analyses (\S \ref{subsec:analyses}) using \approach and \lang. 
\textcolor{\MRColor}{We also present a novel approach that leverages the process-centric analyses through online monitoring of trajectories, and alerts the agents when identified trajectory issues for change of problem-solving strategy (\S \ref{subsub:online-monitoring-analysis}).}

\vspace{-5pt}
\subsection{Process-Centric Metrics}
\label{subsec:metrics}

\approach enables computation of process-centric metrics for analysis of agentic systems. Table~\ref{table:metric-definitions} lists \emph{six}  metrics supported by the current implementation of \approach,\footnote{The rich semantic structure of \approach enables the introduction of an unlimited number of metrics and analyses, similar to CFGs for classic software. This paper discusses the metrics that we believe are more general yet representative.} along with their formal definitions capturing how they can be computed from a given \approach. These metrics capture \emph{the extent of effort} an agent performs to resolve a task. Alone, their values cannot reveal whether the effort is useful or useless. But they can be used to assess how the effort is aligned with the difficulty of the problems agents are trying to solve, or whether agents succeed in solving them. We explain each metric and the intuition about its importance below: \

\begin{table}[t]
    \centering
    \footnotesize
    \caption{Process-centric metrics and their formal definition for \approach $G = (V, TE \uplus SE)$.}
    \vspace{-5pt}
    \label{table:metric-definitions}
    \begin{tabularx}{\linewidth}{l l X@{}}
        \toprule
        \textbf{Metric} & \textbf{Formal Definition} & \textbf{Description}\\
        
        \midrule
        Node Count ($NC$)
        & $|V| = \sum\limits_{v_x \in V} v_x$ 
        & Number of distinct actions.
        \\ 

        \midrule
        Temporal Edge Count ($TEC$)
        & $|TE| = \sum\limits_{e_{x,y} \in TE} e_{x,y}$ 
        & Total number of temporal transitions. 
        \\ 
        

        \midrule
        Loop Count ($LC$)
        & $\sum\limits_{v_m = v_n}\!\{(v_m \to \dots \to v_n) \mid (v_{i-1},v_i)\!\in\!TE\}$ 
        & Number of times the agent decides to repeat a previously executed action.
        \\ 

        \midrule
        Average Loop Length ($ALL)$
        & $\dfrac{\sum\limits_{v_m = v_n}\! \big| \{(v_m \to \dots \to v_n) \mid (v_{i-1},v_i)\!\in\!TE\}\big|}{\sum\limits_{v_m = v_n}\!\{(v_m \to \dots \to v_n) \mid (v_{i-1},v_i)\!\in\!TE\}}$ 
        & The average number of actions involved before an agent decides to repeat a previously executed action.
        \\ 

        \midrule
        Structural Edge Count ($SEC$)
        & $|SE| = \sum\limits_{e'_{x,y} \in SE} e'_{x,y}$ 
        & Total number of structural connections.
        \\ 
        
        \midrule
        Navigation Breadth ($NB$)
        & $\max\limits_{v_x \in V} \big| \{v_y \in V \mid (v_x, v_y) \in SE\} \big|$
        & Maximum out-degree considering structural edges, indicating effort focus.
        \\ 

        
    \bottomrule
\end{tabularx}
\vspace{-5pt}
\end{table}

\begin{itemize}[leftmargin=*]
\renewcommand\labelitemi{-}

    \vspace{-5pt}
    \item \textbf{Node Count ($NC$).}
    Measures the number of distinct actions in the trajectory. A larger $NC$ value implies that the agent needs to execute more diverse actions to solve the problem. From Figure~\ref{fig:graphectory-example}, we can see that \SA\textsubscript{Dev} executes more actions compared to \SA\textsubscript{DSK-V3} ($13$ versus $9$). As we show later, $NC$ is positively correlated with the strength of the backbone LLM (\S \ref{subsub:llm-analysis}) or problem difficulty (\S \ref{subsub:human-analysis}), each of which entails exploration of a wider range of actions. 

    \vspace{2pt}
    \item \textbf{Temporal Edge Count ($TEC$).}
    This metric measures the total number of temporal transitions between actions. A higher $TEC$ reflects a longer execution path to termination. Figure~\ref {fig:graphectory-example} shows a longer execution chain for \SA\textsubscript{Dev} ($14$) compared to \SA\textsubscript{DSK-V3} ($8$). Similar to $NC$, the values of $TEC$ are positively correlated with the problem difficulty, aligning the overall \approach complexity with problem difficulty. 

    \vspace{2pt}
    \item \textbf{Loop Count ($LC$).}
    This metric measures the number of times the agent decides to repeat a previously executed action. As we demonstrate later (\S \ref{sec:rq1}), larger $LC$ can indicate a non-optimized strategy resulting in the agent being stuck by unsuccessful actions, e.g., the self-loop of edge $9$ in Figure~\ref{fig:graphectory-example}-a, or needing to refine a previous reasoning and planning after a series of actions that cannot solve the problem, e.g., back edge $10$ in Figure~\ref{fig:graphectory-example}-a (\S \ref{subsub:eval-strategy-change}). Alternatively, the agent may need to re-execute an action (or series of actions) to solve a complex problem gradually. $LC$ is similar to the Cyclomatic Complexity~\cite{mccabe1976complexity} for classic software, which measures the degree of branching and the number of unique execution paths. In the context of \approach, branching indicates that local reasoning redirects execution flow to a different path.

    \vspace{2pt}
    \item \textbf{Average Loop Length ($ALL$).}
    This metric measures the average number of actions involved before an agent decides to repeat a previously executed action. Intuitively, the longer it takes for an agent to realize its reasoning and corresponding actions are not effective in solving the problem, the weaker its reasoning or ability to gather proper contexts for reasoning is. As our results show, agents that invest in collecting a richer context during localization are less likely to have long loops, i.e., they can quickly determine reasoning inefficiency and change their strategies accordingly (\S \ref{subsub:agent-analysis} and \S \ref{subsub:eval-strategy-change}).

    \vspace{2pt}
    \item \textbf{Structural Edge Count ($SEC$).}
    This metric measures the total number of structural edges in \approach, reflecting the number of structural regions that the agent explores until termination. A higher number of $SEC$ for \SA\textsubscript{DSK-V3} in Figure~\ref {fig:graphectory-example} demonstrates that it took longer for this agent to find the correct edit location, compared to \SA\textsubscript{Dev}. Our results show that stronger LLMs exhibit more exploration of structural regions, collecting more context to increase the likelihood of problem-solving (\S \ref{subsub:llm-analysis}). 

    \vspace{2pt}
    \item \textbf{Structural Breadth ($SB$).}
    This metric measures the maximum out-degree considering structural edges seen in \approach, indicating the navigation effort focus. A higher number for $SB$ indicates the agent's difficulty in converging to the correct code regions for solving the problem. In the example of Figure~\ref {fig:graphectory-example}-b, \SA\textsubscript{DSK-V3} scans two sibling subdirectories ($SB=2$), indicating its subpar reasoning or tool/command usage to localize the bug in the first attempt. As we discuss later, $SB > 1$ can also indicate more dedicated context gathering, which eventually helps the model to find the correct edit location faster (\S \ref{subsub:agent-analysis}).

\end{itemize}

\subsection{Process-Centric Analyses}
\label{subsec:analyses}

We categorize the process-centric semantic analyses of agents' behavior into two categories: Phase Flow Analyses (\S \ref{subsub:phase-flow-anaysis}) and Pattern Detection (\S \ref{subsub:pattern-detection}). \textcolor{\MRColor}{Such analyses can be done offline at the end of the trajectory, or online---with \approach and \lang built from a partial trajectory to a specific step---for real-time monitoring and analysis (\S \ref{subsub:online-monitoring-analysis}).}

\subsubsection{Phase Flow Analyses}
\label{subsub:phase-flow-anaysis}

Phase Flow Analyses aim to study problem-solving strategies of agents independent of low-level actions. Specifically, these analyses take the \approach and \lang as input, and focus on analyzing phase transitions, strategy changes, and shared strategies. 

\vspace{3pt}
\noindent\emph{Phase Transition analysis.}
Phase transition sequence is the abstract representation of an agent's \lang, and an overview of its problem-solving strategy. 

\vspace{3pt}
\noindent \textbf{Definition 3 (Phase Transition Sequence).} Given a \lang $\mathcal{L}(G,\Phi)$, phase transition analysis aggregates consecutive identical phases in \lang, and provides a phase transition sequence as $\mathcal{PTS}(G,\Phi)=\prod\limits_{p_j \neq p_{j+1}} p_j \in \Phi$.

In the example of Figure~\ref{fig:graphectory-example}-a, the phase transition sequence that \SA\textsubscript{Dev} goes through to solve the problem of \texttt{\small{django-10973}} is \emph{LPLPV}. This representation signals that the agent, after attempting to localize the bug and patch it, decides to make another attempt at localization to repair the issue completely, corroborated by the subsequent validation. In addition to analyzing the phase transitions in individual runs, our pipeline can also aggregate them across all trajectories of an agent and present them as a Sankey diagram, revealing the dominant strategic flows (\S \ref{subsub:phase-transition-eval}).

\vspace{5pt}
\noindent\emph{Strategy Change Analysis.} 
Considering a set of $n$ phases, the phase transition sequence may represent up to $n(n-1)$ unique phase transitions. The phase transition sequence \emph{LPLPV} of \SA\textsubscript{Dev} demonstrates three unique phase transitions: L$\to$P (repeated twice), P$\to$L, and P$\to$V. For a phase transition $p_i \to p_{j}$, the transition follows the logical phase flow if $p_{j}$ is the expected immediate successor of $p_i$. In the context of agentic programming systems and $p_i \in \Phi=\{L,P,V\}$, the logical phase flow is $L \to P \to V$ (first localize the bug, next patch, and finally validate). The phase transition may also represent a strategic \emph{shortcut} or \emph{backtrack}. A strategic shortcut happens when agents skip one or multiple logical phases in the phase transition, e.g., $L \to V$. A strategic backtrack happens when $p_{j}$ is a predecessor of $p_i$, e.g., $P \to L$ or $V \to P$. 

The strategy change analysis takes the phase transition sequence as input, determines the number of unique phase transitions, and flags strategic shortcuts or backtracks for investigation. As our results show, the strategic shortcut of $L \to V$ is common in strong models such as Claude 4, indicating the need to review the generated patch and its dependencies to create and execute validation tests (\S \ref{subsub:eval-strategy-change}). Strategic backtracks also occur for two main reasons: agents may need multiple revisits to logical phases to gradually solve a difficult problem, or reasoning inefficiency that requires an agent to refine its local planning and explore a different strategy to solve the problem. The root cause can be automatically detected by examining the \emph{outcome} $o_x$ of the last action in the phase before the transition: if the outcome's binary value (Definition 1) is false, the backtrack is triggered due to reasoning or tool-usage inefficiency. Otherwise, the agent revisits logical phases as a normal problem-solving process (\S \ref{subsub:eval-strategy-change}). In the strategic backtrack of $P \to L$, if the outcome of the last patching action is false, the agent attempts a better localization. However, if it was successful, the agent likely reviews the source code to find or generate validation tests. 

\begin{algorithm}[t]
\caption{Shared Strategy Analysis}
\label{alg:two-stage-lcp}
\small
\algrenewcommand\algorithmiccomment[1]{\hfill$\triangleright$~#1}
\begin{algorithmic}[1]

\Require List of \lang $[\mathcal{L}_i]_{i=1}^n$
\Ensure Longest Common Shared Strategy Pattern $\mathcal{LCP}$

    \State $[\mathcal{PTS}] \gets [\,]$ 
    \ForAll{$\mathcal{L}_i \in [\mathcal{L}_i]_{i=1}^n$}
        \State $\mathcal{PTS}_i \gets$ getPhaseTransitionSequence($\mathcal{L}_i$)
        \State $[\mathcal{PTS}] \gets$ append($[\mathcal{PTS}]$, $\mathcal{PTS}_i$)
    \EndFor
    \State $\pi^\star \gets$ generalizedSequentialPattern($[\mathcal{PTS}_i]_{i=1}^n$)

  \State $B_j \gets [\,]$ for $j \in \{1, \ldots, |\pi^\star|\}$ \Comment{Initialize buckets to collect phase durations}
  
  \ForAll{$\mathcal{L}_i \in [\mathcal{L}_i]_{i=1}^n$}
    \State $(match,\mathbf{k}) \gets$ findMatch($\pi^\star$,$\mathcal{L}_i$) \Comment{Find the inception of $\pi^\star$ in the \lang}
    \If{match}
      \For{$j = 1$ \textbf{to} $|\pi^\star|$}
        \State $B_j \gets$ append($B_j$,phaseSize($\mathcal{L}_i.\pi(v_{\mathbf{k}_j})$))
      \EndFor
    \EndIf
  \EndFor
  \State $\mathbf{RL} \gets [\min B_j \text{ if } B_j \neq \emptyset \text{ else } 1]_{j=1}^{|\pi^\star|}$ \Comment{Minimum run-length bounds}

\State $\mathcal{LCP} \gets$ getLCP($\pi^\star$,$RL$)
\State \textbf{return} $\mathcal{LCP}$
\end{algorithmic}
\end{algorithm}


\vspace{5pt}
 \noindent\emph{Plan Compliance/Violation Analysis.}
 This analysis evaluates whether an agent follows the \emph{suggested} problem-solving plan during execution. This plan is usually provided to agents in natural language in the system prompts that initiate the agents~\cite{SWE-agent-plan,open-hands-plan}, and can be formally represented as a \emph{golden phase sequence (grammar)} $\mathcal{L}^\star \in \Phi^{*}$ (e.g., $\mathcal{L}^\star = L_{*} P_{*} V_{*}$ for SWE-agent, indicating localization, patching, and validation in order). A trajectory complies with the plan if its \lang $\mathcal{L}(G,\Phi)$ respects the ordering of $\mathcal{L}^\star$, independent of the specific run-length subscripts; phase revisits are allowed to support iterative problem solving on harder instances. Skipped or reordered phases are recorded as plan violations. For example, the trajectory of \SA\textsubscript{Dev} (Figure~\ref{fig:graphectory-example}-a) with $\mathcal{L} = L_{5} P_{5} L P V$ complies with the plan. In contrast, trajectories with $\mathcal{L} = PLV$ or $\mathcal{L} = LPL$ violate the plan due to phase reordering (patching before localization) and skipping the validation phase, respectively.

\vspace{15pt}
\noindent\emph{Shared Strategy Analysis.}
The last step of the Phase Flow Analyses (shown in Algorithm~\ref{alg:two-stage-lcp}) extracts shared strategic subsequences for a given agent across different trajectories, obtained by executing actions to solve different problems. Taking a list of \lang instances as input, Algorithm~\ref{alg:two-stage-lcp} first computes their corresponding phase transition sequences (lines $1$--$3$), and then uses the Generalized Sequential Pattern (GSP) algorithm~\cite{gsp1996} to find the longest common subsequences across all phase transition sequences (line $4$).\footnote{We set \texttt{min\_support} to 0.3 in the GSP algorithm for our experiments. This threshold ensures statistical robustness while capturing meaningful behavioral patterns, consistent with established pattern mining practices~\cite{min_sup}. 
} At the next step, it finds the inception of the identified common pattern in each \lang (line $7$) to determine phase lengths corresponding to the run-length encoding (lines $8$--$10$). To account for the difference in phase lengths, Algorithm~\ref{alg:two-stage-lcp} takes the minimum run-length across all \lang (line $11$) to adjust the common strategy pattern (line $12$).

Consider the two agents whose \approach is shown in Figure~\ref{fig:graphectory-example}. Given the two \lang of these agents, $L_5P_5LPV$ and $L_6P$, and their phase transition sequences, $LPLPV$ and $LP$, Algorithm~\ref{alg:two-stage-lcp} determines $LP$ as the longest common pattern. Next, it checks the run-length of the common pattern in the \lang, and returns $L_5P$ as the shared strategic pattern between these two agents. As we will see (\S \ref{subsub:eval-shared-strategy}), the common problem-solving strategies in agents vary based on problem difficulty, whether or not the problem is solved successfully, and the strength of the backbone LLM. 

\subsubsection{Pattern Detection}
\label{subsub:pattern-detection}

\approach and \lang are rich structures that enhance the mining of both \emph{known} and \emph{unknown but common} patterns. Phase Flow Analysis, and specifically Shared Strategy Analysis, demonstrates the usefulness of the structures in mining common unknown patterns. Pattern Detection Analysis complements that by enabling the search for known patterns in \approach or \lang of agents. 

The prominent use case of pattern detection is finding known failure modes or suspected strategic inefficiencies in \approach. As we show later, this enables a systematic and large-scale analysis of programming agents' failure modes (\S \ref{sec:rq3}): users can sample a relatively small number of failures, manually investigate them to determine a list of (anti-)patterns with respect to the sampled data, and then search for these patterns among other runs systematically and without manual effort. 

\subsection{Online Monitoring and Intervention}
\label{subsub:online-monitoring-analysis}

\textcolor{\MRColor}{
Process-centric analysis, when performed postmortem at the end of the trajectory, can determine overall issues in the trajectory, such as inefficiency patterns and plan violations. Online process-centric analyses, on the other hand, if followed by proper interventions, can detect trajectory issues \emph{early} (or even before they manifest) and \emph{prevent} them from impacting the overall trajectory and final outcome.
Online process-centric analysis requires incremental construction of \approach and \lang throughout the trajectory. At each step, the online monitor will update the partial \approach and \lang induced by the current trajectory. It then analyzes them for (1) plan compliance and (2) checks for potential inefficiencies or other failure modes, if specified. We identify a series of heuristics that can \emph{potentially} (but not necessarily) indicate an inefficient or regressive behavior. These heuristics are\footnote{The heuristics are orthogonal to our design, i.e., more heuristics or even specific inefficiency patterns can be added later.}:
}

    \begin{itemize}[leftmargin=*]
        \item \textbf{H1.} A large loop, i.e., a series of consecutive actions connected by a back edge, indicates that an agent performed a series of actions that did not solve the issue, requiring a change of strategy. 
        
        \item \textbf{H2.} A misalignment between the temporal and structural edges indicates that an agent moved back and forth within the project structure rather than exploring steadily.  
        
        \item \textbf{H3.} A long phase length indicates that the agent attempted a series of actions within the same phase without making progress.
        
        \item \textbf{H4.} A failure outcome for any node shows potential inefficiencies in reasoning or tool usage. 
        
    \end{itemize}

\textcolor{\MRColor}{
When the analysis of partial \approach and \lang determines the existence of inefficiency signals or plan violations in the current trajectory, the online monitor will notify the agent with diagnostic messages, e.g., "You may be stuck in localization. Focus on relevant code and run tests to reproduce and localize the bug if necessary." Depending on the nature of the issue, interventions may block the most recent action that triggered the intervention and revert the trajectory, \approach, and \lang to the previous step. For example, consider an agent that submits the patch after localization \emph{without} any validation, resulting in a plan violation; the intervention component blocks the most recent action, \texttt{\small{submit}}, that violated the plan, giving the agent an opportunity to refine the strategy. For inefficiency signals, intervention involves no rollback, since the history of taken actions may be required for reasoning about the steps given the diagnostic message. This design enables \approach and \lang to serve as a general interface for online monitoring, debugging, and control of agentic systems.}

\section{Experiments}
\label{sec:rqs}

We demonstrate the usefulness of \approach and \lang in enabling systematic analysis of agentic programming systems through the following research questions:

\begin{enumerate}[label=\bfseries RQ\arabic*.]

    \item \textbf{Process-centric Metrics.} What are the \approach characteristics of programming agents concerning process-centric metrics? What factors impact the metric values? 

    \item \textbf{Analysis of Problem-solving Strategies.} What are the most common problem-solving strategies in programming agents? What factors impact their strategic behavior? To what extent \lang of agents follows the expected planning? To what extent do agents evolve or change their problem-solving strategies during execution?

    \item \textbf{Inefficiency Pattern Analysis.} What are the common \approach (anti-)patterns consistent among the agents' behavior? How prevalent are these patterns?

    \item \textbf{\textcolor{\MRColor}{Online Monitoring and Intervention.}} 
    \textcolor{\MRColor}{Can online process-centric analysis and intervention mitigate trajectory-level issues and improve resolution rate?}

\end{enumerate}

\subsection{Experiment Setup}
\label{subsec:setup}

We study two autonomous, open-process, and widely used programming agents: \SA (SA) and \OH (OH) for the evaluation.\footnote{We do not study pipelines such as Agentless~\cite{xia2024agentless}, as the defined pipeline denotes the agent's strategies. We also do not include Refact~\cite{refact} due to known persistent technical issues during setup and execution.} We follow each framework’s default configurations to run the experiments: \SA uses a per-instance cost cap of \$2; \OH runs for a maximum of $100$ trajectory iterations; both use their default file-viewing and code-editing tools. We pair the agents with four LLMs: \textbf{DeepSeek-V3} 671B MoE (DSK-V3)~\cite{deepseek-v3}, \textbf{DeepSeek-R1} (DSK-R1)~\citep{deepseek-r1}, \textbf{Devstral-small-2505} 24B coding-specialized open weights (Dev)~\cite{devstral-small}, and \textbf{Claude Sonnet 4} (CLD-4)~\cite{claude-sonnet-4}. Our choice of LLMs includes both general and reasoning models, as well as frontier and open-source models, providing \emph{eight} $\langle$agent, model$\rangle$ settings and enabling analysis of results concerning different training strategies. 

\begin{figure}[t]
\includegraphics[width=0.9\columnwidth]{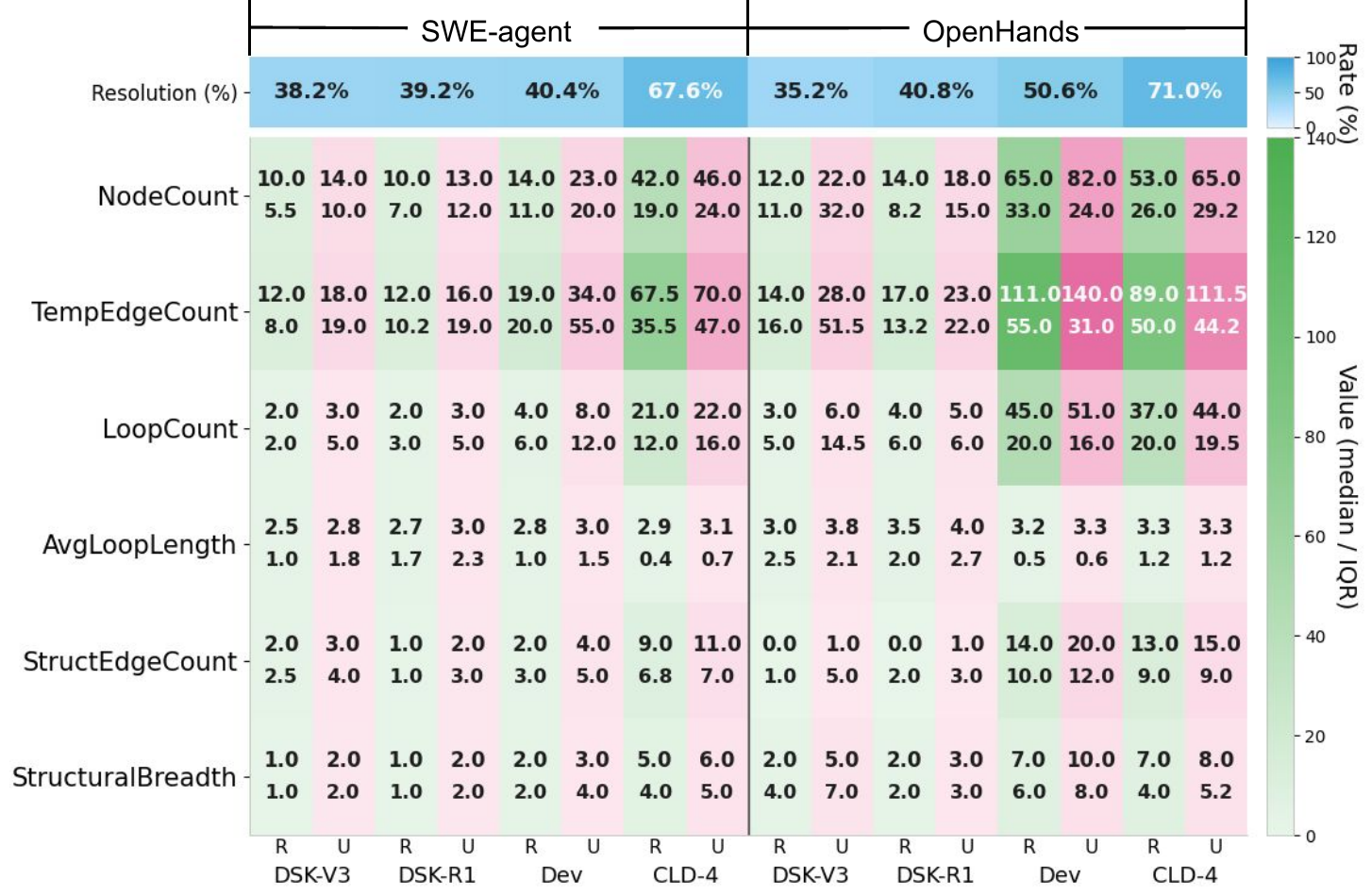}
\vspace{-10pt}
    \caption{Process-centric metrics by agent–model pair. Columns are grouped by \emph{agent} and \emph{model}; within each pair, the left column \colorbox[HTML]{95D0B3}{R} summarizes \emph{resolved} runs and the right column \colorbox[HTML]{EE91BB}{U} summarizes \emph{unresolved} runs. The top blue row reports the resolution rate. Each heatmap cell reports $\left\langle \substack{Median \\ IQR} \right\rangle$ for the metric in that row}
    \label{fig:trajectory_heatmap}
\end{figure}

We report \emph{Pass@1} results, i.e., the first submission per issue. LLMs are inherently non-deterministic, which can impact their trajectories and outcomes over multiple runs. We account for this inherent non-determinism by performing cross-analysis of the trajectories over multiple problems with different levels of difficulty, reporting the median and interquartile range values, and analyzing the aggregated results over all problems in addition to individual analysis of $\langle$agent, model$\rangle$ pairs. All experiments use \textbf{SWE-Bench Verified}~\cite{swebench-verified}, which consists of $500$ real GitHub issues (a human-validated subset of SWE-Bench), providing a total of $4000=500\times8$ trajectories for analysis.

\subsection{RQ1: Process-Centric Metrics}
\label{sec:rq1}

In this research question, we compute the values of process-centric metrics (Table~\ref{table:metric-definitions} in \S \ref{subsec:metrics}) for each $\langle$agent, model$\rangle$ pair. Figure~\ref{fig:trajectory_heatmap} shows the collected values across resolved (green columns marked with R) and unresolved (pink columns marked with U) issues. The top blue row reports the \emph{resolution rate}, and each heatmap cell shows the \emph{median} of the metric and \emph{Interquartile Range (IQR)}, representing the spread around the median. Darker colors correspond to higher metric values.

\subsubsection{Analysis across Agents.}
\label{subsub:agent-analysis}
Overall, we can observe that the \approach of \OH are more complex compared to that of \SA, corroborated by the higher values of metrics. This indicates that \OH puts more effort into solving the problem, likely due to the detailed instructions in its system prompt~\cite{open-hands-plan} compared to \SA~\cite{SWE-agent-plan} (detailed discussion in \S \ref{sec:design}). 

\begin{figure}[t]
\includegraphics[width=0.9\linewidth]{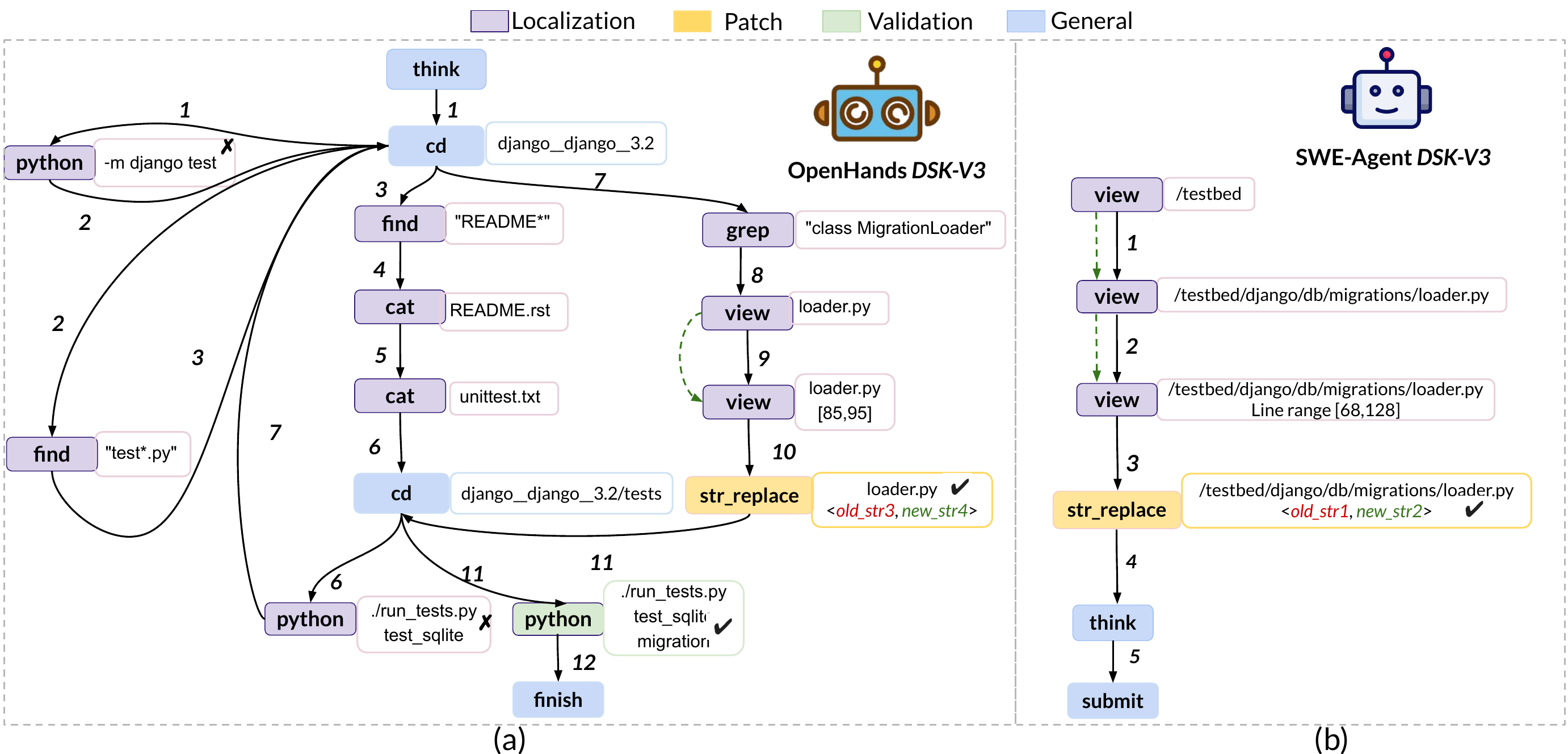}
\vspace{-8pt}
    \caption{\approach of \OH\textsubscript{DSK-V3} (a) and \SA\textsubscript{DSK-V3} (b) for problem \texttt{\small{django-13820}}}
    \label{fig:agent_analysis}
\end{figure}

Figure~\ref{fig:agent_analysis} shows the \approach of \OH\textsubscript{DSK-V3} (left) and \SA\textsubscript{DSK-V3} (right) for solving SWE-bench issue \texttt{\small{django-13820}}. Both agents resolve the issue. However, the \approach of \OH\textsubscript{DSK-V3} is more complex compared to \SA\textsubscript{DSK-V3}. 
Looking more closely, \OH\textsubscript{DSK-V3} first runs a test to localize the issue, searches and inspects related files, executes additional tests to ensure proper localization (step $7$), then navigates back to the source to read the buggy code and generate the patch. It finally runs tests to validate the fix. \OH also issues composite shell commands within single steps (e.g., \texttt{\small{cd}} \&\& \texttt{\small{python}} chained), showing denser reasoning per step and more frequent navigation switches.
In contrast, the \approach of \SA\textsubscript{DSK-V3} (Figure~\ref{fig:agent_analysis}-b) is simple: it narrows quickly to the target file via a few views and applies a single edit (step $4$) and then submits that patch with no test execution for validation. The two behaviors illustrate a trade-off: both solve the problem, \OH invests extra effort to gather context and validate, whereas \SA relies on concise localization and a direct edit path.

\subsubsection{Analysis across LLMs.}
\label{subsub:llm-analysis}
Stronger LLMs, e.g., Claude Sonnet 4, which consistently ranks among the top-performing LLMs on different tasks, tend to have a more complex \approach. As we will discuss in more detail (\S \ref{sec:rq2}), this is because they tend to gather more context and run extra tests to ensure the patch works and avoid regressions. The Devstral model, which is a relatively small open-source model, also has a more complex \approach, specifically when used with \OH. We believe this is because it has been trained using the \OH agent scaffold for software tasks~\cite{devstral-intro}, encouraging extra searches and test runs before committing edits. 

\subsubsection{Analysis across Repair Status.}
\label{subsub:repair-status-analysis}
We further want to see if there is a correlation between the process-centric metrics and repair status, i.e., whether the agent can resolve the issue. To that end, we perform the Mann-Whitney U test~\cite{mann1947test} for each metric across all $\langle$agent, model$\rangle$ pairs. This non-parametric statistical test ranks all observations from both groups (a process-centric metric and repair status) and compares the sum of ranks between the two groups. A significant result indicates that the distributions of the two groups differ. 

\begin{figure}[t]
\includegraphics[width=0.9\columnwidth]{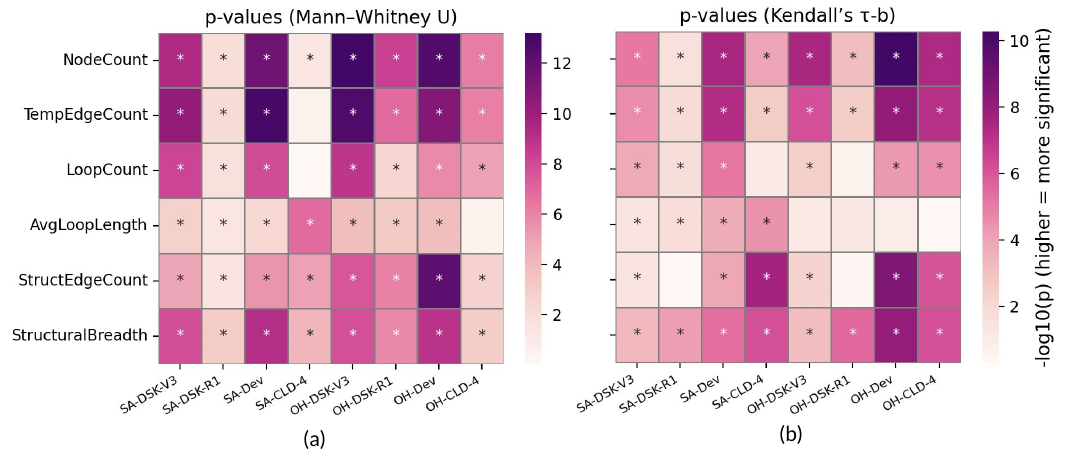}
\vspace{-8pt}
    \caption{\textit{p}-values of statistical tests on (a) issue repair status and (b) human difficulty alignment. Cells with $\star$ indicates significant difference ($p\leq 0.05$)}
    \label{fig:sig-test}
\end{figure}

Figure~\ref{fig:sig-test}-a shows the \textit{p}-values: most cells show significant differences (\textit{p} $\leq 0.05$, marked with~$*$), indicating that the process-centric metrics can distinguish successful and unsuccessful repairs, except for Claude as the backbone LLM. As we discuss later (\S \ref{sec:rq2}--\S \ref{sec:rq3}), other LLMs prefer shortcuts in structural navigation, i.e., reasoning and jumping to the suspected file location without broad exploration. On the other hand, Claude 4 tends to visit more structural regions to gather the context needed for effective problem-solving. Similarly, other LLMs terminate with shorter trajectories as soon as they resolve the issue. In contrast, Claude Sonnet 4 consistently validates the generated patches with multiple rounds of test execution, resulting in a more complex \approach.

When comparing a reasoning model, DeepSeek-R1, with its general counterpart, DeepSeek-V3, we see that the correlation between process-centric metrics and repair status is overall positive; yet, the \emph{p}-values for DeepSeek-R1 are smaller. Through manual investigation of the DeepSeek-R1 trajectories, we observed many early terminations due to runtime issues caused by its failure to provide model responses in the correct format, which is a known issue.\footnote{https://aider.chat/2024/08/14/code-in-json.html}

\subsubsection{Analysis across Problem Difficulty.}
\label{subsub:human-analysis}

Finally, we assess, using the process-centric metrics, how problem difficulty impacts agent behavior. SWE-Bench Verified determines problem difficulty based on how long it takes a human to fix the issues: \textit{Easy} (less than $15$ minutes), \textit{Medium} ($15$--$60$ minutes), \textit{Hard} ($1$--$4$ hours), and \textit{Very Hard} (more than $4$ hours). For this analysis, we used Kendall’s $\tau_b$ test~\cite{kendall1938}. Figure~\ref{fig:sig-test}-b shows the \textit{p}-values for all agent–model pairs, where most cells show significant positive trends ($p\leq0.05$, marked with~$*$). i.e., a consistent monotonic trend between each metric and the difficulty ranking. This analysis demonstrates that, similar to humans who need more time to fix more difficult problems, agents also put more effort into resolving more difficult issues. 

\begin{figure}[t]
\includegraphics[width=0.95\linewidth]{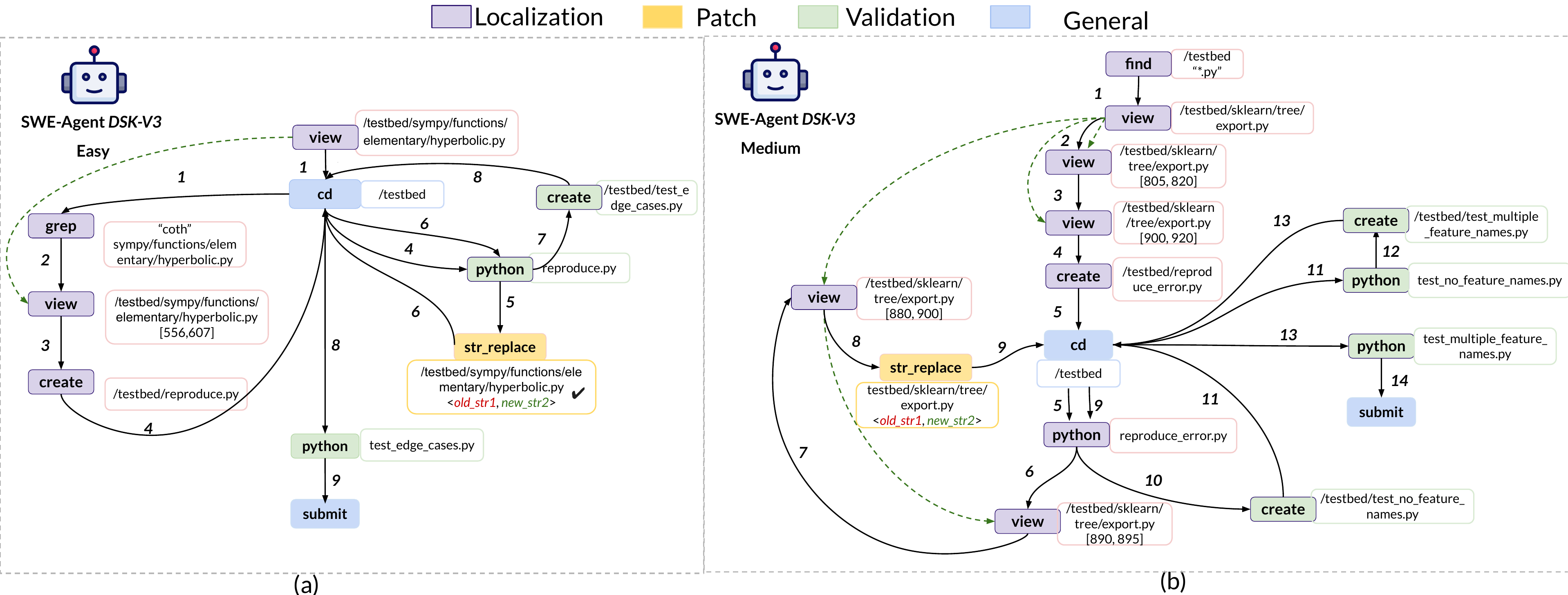}
\vspace{-8pt}
    \caption{\approach of \SA\textsubscript{DSK-V3} for (a) problem \texttt{\small{sympy-13480}} (easy) and (b) problem \texttt{\small{scikit-learn-14053}} (medium)}
    \vspace{-5pt}
    \label{fig:difficulty_example}
\end{figure}

Figure~\ref{fig:difficulty_example} contrasts two resolved cases for \SA\textsubscript{DSK-V3}. The \textit{Easy} task (Figure ~\ref{fig:difficulty_example}-a, \texttt{\small{sympy-13480}}) converges in 10 steps: the agent jumps straight to the buggy file, creates and runs a small reproduction test, edits, re-runs the test, and submits. The \textit{Medium} task (Figure~\ref{fig:difficulty_example}-b, \texttt{\small{scikit-learn-14053}}) takes 15 steps: the agent first searches the directory and scrolls up and down to locate the buggy area. It also adds extra tests to cover edge cases before submitting. Together with the statistics, these examples show a clear alignment: as human difficulty increases, agents require greater effort to find the correct solution (e.g., perform more extensive searches and include more tests).

\vspace{-5pt}
\begin{findingsbox}
\textbf{Findings.} Process-centric metrics correlate with both repair success and human-rated difficulty. Successful runs tend to be shorter and focused, while unresolved ones exhibit longer paths and redundant loops. The \approach complexity is also affected by agent design and backbone LLMs. Stronger models produce a denser \approach with broader exploration.
\end{findingsbox}

\subsection{RQ2: Analysis of Problem-Solving Strategies}
\label{sec:rq2}

Beyond whether agents ultimately succeed or fail in repairing an issue, we further investigate the underlying agents’ behavior using the Phase Flow Analysis series (\S \ref{subsub:phase-flow-anaysis}). 

\subsubsection{Phase Transitions Analysis}
\label{subsub:phase-transition-eval}

\begin{figure}[t]
  \centering
  \includegraphics[width=0.9\linewidth]{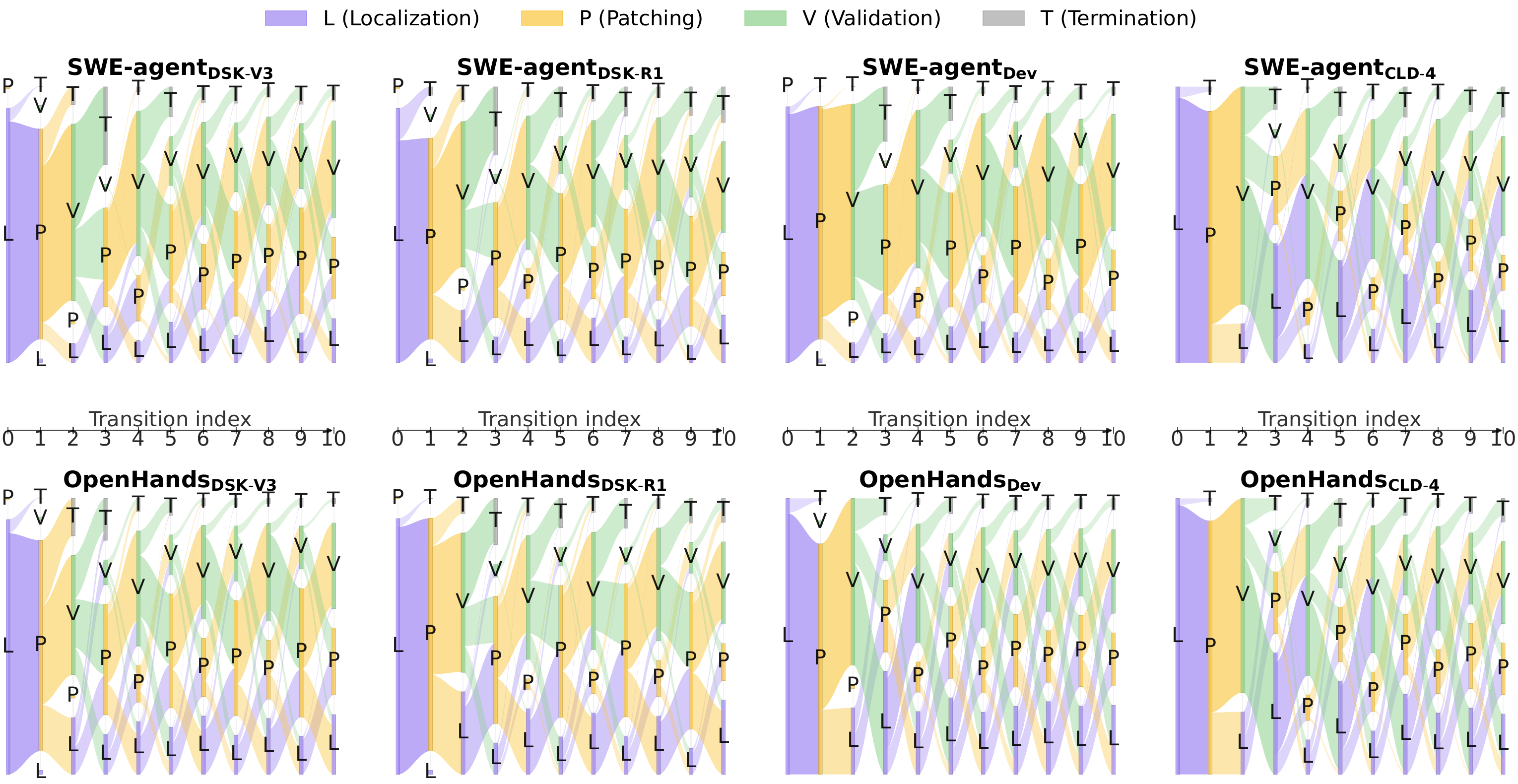}
  \vspace{-8pt}
  \caption{Phase transition sequences (cut-off = $10$; \textcolor{\MRColor}{bar height = trajectory counts in each phase at this step; flow thickness = the number of trajectories that transition from one phase to another})}
  \vspace{-5pt}
  \label{fig:sankey-phases}
\end{figure}

\begin{figure}[t]
   \centering
   \includegraphics[width=1\linewidth]
   {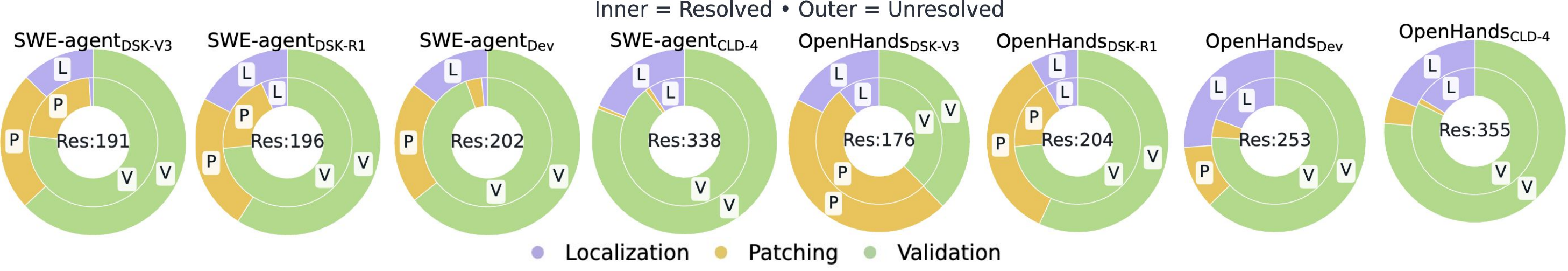}
   \vspace{-15pt}
   \caption{Distribution of terminal trajectory phases}
   \vspace{-10pt}
   \label{fig:end-phases}
 \end{figure}

To better illustrate the phase flow of the studied $\langle$agent, model$\rangle$ pairs, we present the aggregated phase transition sequence ($\mathcal{PTS}$) of each pair across the $500$ SWE-bench Verified issues for the first $10$ phase transitions (Figure~\ref{fig:sankey-phases}). The \textcolor{\MRColor}{average length of Phase Transition Sequences} among the studied agents is $8.39$, making the cut-off of $10$ reasonable to reflect overall phase transition flow. 
We also study the termination phases of $\mathcal{PTS}$s separately in Figure~\ref{fig:end-phases}, covering cases with more than $10$ phase changes (max = $188$).

Figure~\ref{fig:sankey-phases} shows that the majority of $\mathcal{PTS}$s across all $\langle$agent, model$\rangle$ pairs \emph{start} with localization (\(L\)), reflecting an initial effort to localize the bug.
Using the same LLM, \SA usually leaves localization \(L\) within one or two transitions (indicated by the size of $L$ bars at each transition index) and quickly settles into short \((P,V)\) cycles, indicating an early attempt to generate and validate a patch. In contrast, \OH tends to remain in \(L\) for longer during the first ten iterations, suggesting a longer localization process. This can also explain observing \SA terminating sooner than \OH, corroborated by higher flow density to $T$ in earlier transition indices. 

Figure~\ref{fig:end-phases} illustrates the terminal phase, categorized by agents' success in repairing the problems (inner and outer donuts reflect the distribution of the terminal phase for resolved and unresolved issues, respectively). From this figure, we see that resolved trajectories almost always conclude in \(V\), reflecting successful validation and convergence on a good patch. Unresolved runs, however, frequently end in \(L\) or \(P\), which means the fixing process is incomplete; the agent either tries multiple rounds of bug localization or oscillates between generating patches without validation. 

\subsubsection{Strategy Change Analysis}
\label{subsub:eval-strategy-change}

\begin{figure}[t]
  \centering
  \includegraphics[width=0.88\linewidth]{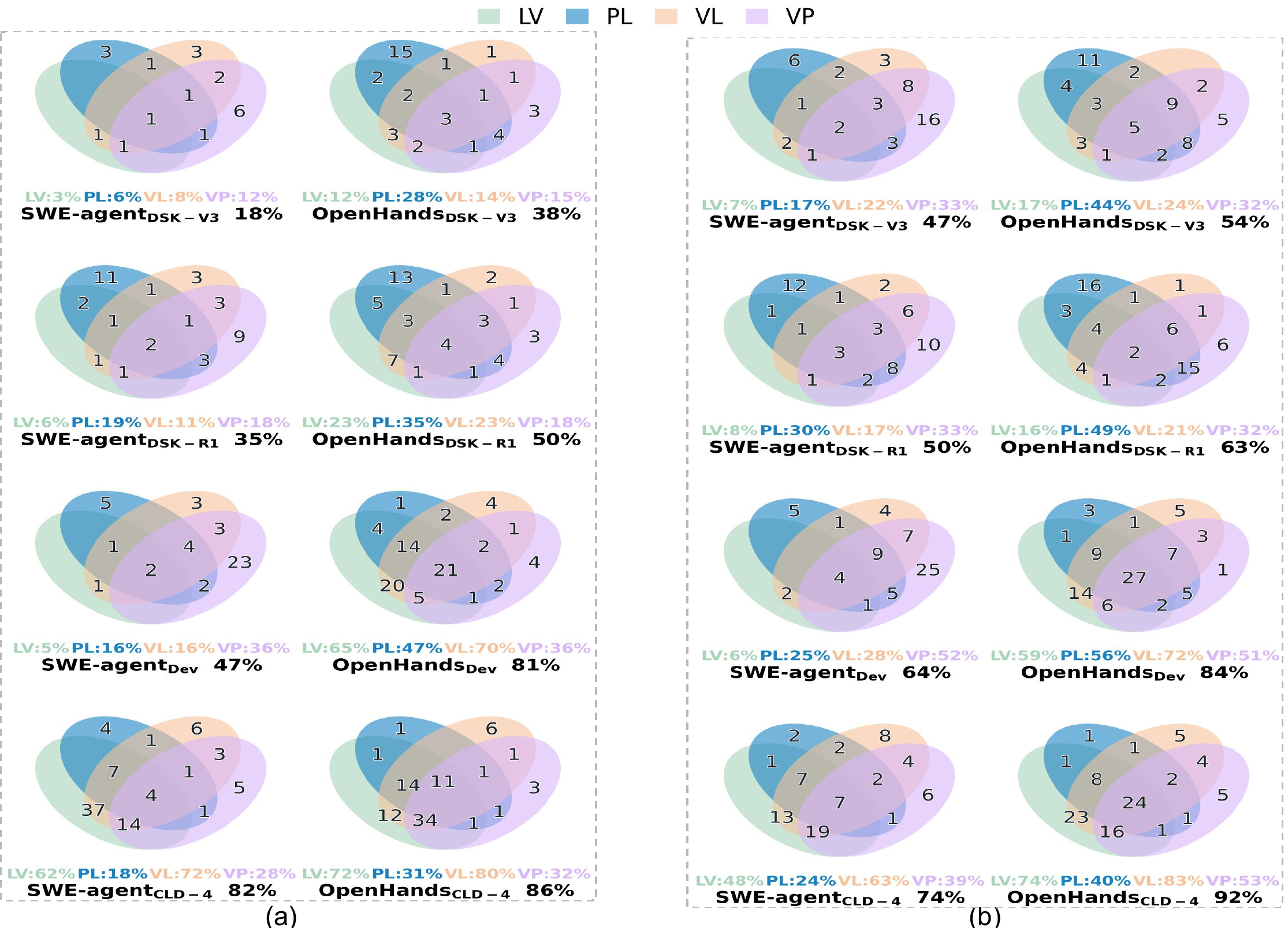}
  \vspace{-10pt}
  \caption{\textcolor{\MRColor}{Phase change distribution across agents and models: (a) Resolved and (b) Unresolved}}
  \vspace{-5pt}
  \label{fig:phase_change_venn}
\end{figure}

As discussed in \S \ref{subsub:phase-flow-anaysis}, and we observe from the phase transition sequences of Figure~\ref{fig:sankey-phases}, agents frequently change phases (and hence, their strategies) to accomplish a task. We focus our analysis on the strategic shortcut ($L \to V$) and backtrack ($P \to L$, $V \to P$, and $V \to L$). Figure~\ref{fig:phase_change_venn} shows the number of instances per each $\langle$agent, model$\rangle$ pair that their $\mathcal{PTS}$ has one of the strategic shortcuts or backtracks. The percentages below the Venn diagrams show the percentage of instances that had at least one of the studied strategy changes.\footnote{Note that the denominator for the diagrams in Figure~\ref{fig:phase_change_venn}-a and Figure~\ref{fig:phase_change_venn}-b are different, depending on the resolved instances for each $\langle$agent, model$\rangle$ pair.}
From this figure, we can see that stronger pairs (\SA/\OH with Claude Sonnet 4 or \OH with Devstral) exhibit a significantly higher number of $L \leftrightarrow V$ chains than others, demonstrating their extra effort on validation and context gathering before committing new edits. Unresolved instances exhibit higher rates and overlaps across all patterns, indicating a greater need for strategy switching to address potential failures. 

\begin{figure}[t]
  \centering
  \includegraphics[width=0.8\linewidth]{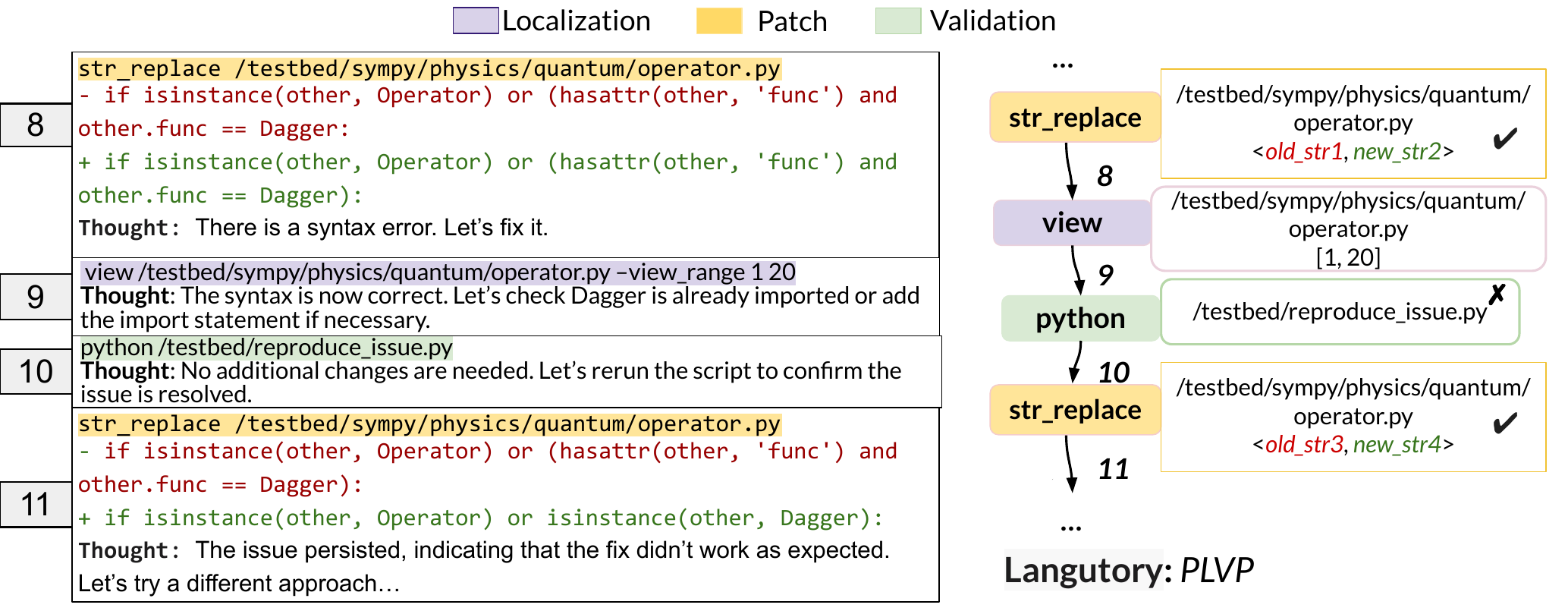}
  \vspace{-8pt}
  \caption{Strategy changes of \SA\textsubscript{DSK-V3} while attempting to resolve \texttt{\small{sympy-19783}} problem}
  \label{fig:phase_change_example}
\end{figure}

Figure~\ref{fig:phase_change_example} shows a snapshot of \SA\textsubscript{DSK-V3} trajectory (left side) and \approach (right side), trying to resolve the issue \texttt{\small{sympy\_\_sympy-19783}} of SWE-bench. Our Strategy Change Analysis flagged this instance with one shortcut strategy ($L \to V$) and two backtrack strategies ($P \to L$ and $V \to $P). Based on the outcome analysis, the second strategic backtrack was marked as an indicator of reasoning inefficiency (\S \ref{subsub:phase-flow-anaysis}). Looking at the thought process of the agent confirms this analysis: this agent goes through multiple backtracking and shortcuts to ensure the correctness of the solution. At trajectory step $8$, the agent modifies the code to fix a syntax error ($P$). At the next step, instead of executing tests for validation, the agent reviews the code again to determine if there are additional bugs ($P \to L$). Then, it reruns the previously created reproduction script to validate the patch ($L \to V$). Test execution reveals that the bug persists, and the agent makes another repair attempt to fix the bug ($V \to P$). The last strategic backtrack is indeed due to an inefficient reasoning at step $8$, suggesting a wrong edit that did not fix the bug. \SA\textsubscript{DSK-V3} ultimately fixes this issue, but at the cost of additional trajectory steps. 

\subsubsection{Shared Strategy Analysis}
\label{subsub:eval-shared-strategy}

Table~\ref{tab:rq2-lcp} shows the top $\mathcal{LCP}$s of studied <agent,model> pairs across difficulty levels and repair status. The \texttt{\small{min\_support}} of $0.3$ in GSP algorithm detects all the longest common patterns persistent among at least $30\%$ of the instances. Table~\ref{tab:rq2-lcp} shows the most prevalent longest common patterns,
which represent the dominant problem-solving strategy. Overall, \emph{resolved instances of easy problems} exhibit structured and well-ordered phase transitions, typically following concise $\langle L,P,V \rangle$ cycles that reflect a disciplined pipeline of localization, patch generation, and validation. As task difficulty increases, these sequences extend moderately (e.g., $\langle L_6,P,V,L,V \rangle$), indicating deeper, yet still coherent strategy chains.
In contrast, \emph{unresolved} trajectories show greater redundancy and longer $(P,V)$ or $(L,P)$ repetitions, such as $\langle L_3,(P,V)^4 \rangle$ for \SA\textsubscript{Dev}. Across backbone models within each agent, stronger models like Claude Sonnet 4 demonstrate broader phase exploration with all three phases in longer and more diverse sequences, while weaker ones (e.g., DeepSeek-V3) tend to produce shorter, less exploratory flows, which may limit their ability to handle more complex cases and result in lower repair rates in the end.

\makeatletter
\newcommand{\pat}[1]{%
  \ensuremath{\langle #1 \rangle}%
}
\makeatother
\newcolumntype{Z}{>{\raggedright\arraybackslash}X}
\setlength{\tabcolsep}{3pt}
\renewcommand{\arraystretch}{1.06}

\begin{table}[t]
\centering
\footnotesize
\caption{Top $\mathcal{LCP}$s of studied $\langle$agent, model$\rangle$ pairs, conditioned on repair status and problem difficulty. Phases: L=\colorbox[HTML]{D4CCE6}{Localization}, P=\colorbox[HTML]{FFE18E}{Patching}, V=\colorbox[HTML]{D4E7CD}{Validation}. Superscripts denote pattern repetition and subscripts denote phase duration (run-length from \lang).}
\label{tab:rq2-lcp}
\begin{adjustbox}{max width=\linewidth}
\begin{tabularx}{\linewidth}{lllZZZZ}
\toprule
\textbf{Agent} & \textbf{Status} & \textbf{Difficulty} & \textbf{Dev} & \textbf{DSK-V3} & \textbf{DSK-R1} & \textbf{CLD-4} \\
\midrule
\multirow{6}{*}{\textbf{SWE-agent}}
& \multirow{3}{*}{\textbf{Resolved}}
& easy      & \pat{L_2, P, V} & \pat{L, P, V} & \pat{L, P, V} & \pat{L, P, (V, L)^2} \\
& & medium  & \pat{L, (P, V)^2} & \pat{L, P, V} & \pat{L, P, V} & \pat{L_6, P, V, L, V} \\
& & hard    & \pat{(P, V_6)^2, L_{24}} & \pat{P, V_4} & \pat{L_3, P} & \pat{P, (V, L)^2, V_2} \\
\cmidrule(lr){2-7}
& \multirow{3}{*}{\textbf{Unresolved}}
& easy      & \pat{L_5, (P, V)^2} & \pat{L, (P, V)^2} & \pat{L, P, V} & \pat{L_3, P, V, L, V} \\
& & medium  & \pat{L_2, (P, V)^3} & \pat{L, (P, V)^2} & \pat{L, (P, V)^2} & \pat{L_7, P, V, L, V} \\
& & hard    & \pat{L_3, (P, V)^4} & \pat{L_3, (P, V)^2} & \pat{L, (P, V)^2} & \pat{L_4, P, V, L, V} \\
\midrule
\multirow{6}{*}{\textbf{OpenHands}}
& \multirow{3}{*}{\textbf{Resolved}}
& easy      & \pat{L_8, P, V, L, V} & \pat{L, P, V} & \pat{L, P, V} & \pat{L_4, P, (V, L)^2} \\
& & medium  & \pat{L_{10}, P, V, L, V} & \pat{L_2, P, V} & \pat{L, P, V} & \pat{L_3, P, (V, L)^2} \\
& & hard    & \pat{P, L, (P, V)^2} & \pat{(L_8, P)^2, (V, L)^2} & \pat{P_2, (L_2, P, V_2)^2} & \pat{V, L_2, (P, V)^2} \\
\cmidrule(lr){2-7}
& \multirow{3}{*}{\textbf{Unresolved}}
& easy      & \pat{L_{18}, P, V, L, V} & \pat{L_3, (P, V)^2} & \pat{(L_2, P)^2} & \pat{L_8, P, (V, L)^2} \\
& & medium  & \pat{L_8, P, V, L, V} & \pat{L_2, P, V} & \pat{(L_3, P)^2, V} & \pat{L_5, P, V, L, V} \\
& & hard    & \pat{(L_{27}, P)^3} & \pat{(L_2, P)^2} & \pat{(L, P)^2} & \pat{P, (V, L)^2, V} \\
\bottomrule
\end{tabularx}
\end{adjustbox}
\end{table}

\begin{findingsbox}
\textbf{Findings.} Successful runs typically follow structured localization, patching, and validation flows that align with the agents’ intended plans. As task difficulty increases, the resolved trajectories extend but remain coherent, while unresolved ones exhibit repetitive or disordered phase transitions. Stronger LLMs demonstrate richer and more adaptive strategies.
\end{findingsbox}

\subsection{RQ3: Inefficiency Pattern Analysis}
\label{sec:rq3}

\textcolor{\MRColor}{
Unlike prior work \cite{liu2025empirical} that heavily relies on manual analysis of trajectories, the rich structure of \approach enables us to mine meaningful patterns systematically (\S \ref{subsub:pattern-detection}). We are interested in identifying \emph{anti-patterns} that reflect inefficient or regressive behavior. To this end, we use the inefficiency heuristics introduced in \S \ref{subsub:online-monitoring-analysis} and examine trajectories for their existence\footnote{We make no claim that our heuristics, and hence, identified anti-patterns are complete. However, this study serves as one of a kind to systematically analyze agents' trajectories for inefficiency patterns, motivating future analyses.}. We sample $15\%$ of the studied \approach, \emph{equally} from resolved and unresolved instances for each <agent,model> pair, accounting for project diversity to avoid sampling bias. This provides us with $600$ samples across agents and models, ensuring that patterns we find in these samples persist in $>=10\%$ of all studied trajectories, with $95\%$ confidence interval. We automatically flag sampled instances that contain at least one of the mentioned heuristic patterns. Then, we manually review the thoughts, actions, and observations at each step to identify anti-patterns. Finally, we systematically search for the anti-patterns in all $4000$ \approach and report prevalence.
}

\begin{table}[t]
\centering
\footnotesize
\caption{Identified anti-patterns from the manual analysis of sampled \approach. The first four rows belong to the \colorbox[HTML]{D4CCE6}{Localization} phase, and the last five rows belong to the \colorbox[HTML]{FFE18E}{Patching} phase.}
\vspace{-8pt}
\label{tab:common-patterns}
\renewcommand{\arraystretch}{1.15}
\begin{tabular}{p{0.03\linewidth}p{0.17\linewidth}p{0.54\linewidth}p{0.18\linewidth}}
\toprule
\textbf{ID} & \textbf{Pattern} & \textbf{Description} & \textbf{Heuristic Patterns} \\
\midrule
\textbf{RV} & \textsc{RepeatedView} &
Agent revisits the same structural region & $H1$, $H3$ \\

\textbf{ZO} & \textsc{ZoomOut} &
The temporal order of view actions contradicts the structural hierarchy from a deeper to a shallower level in the project structure & $H2$, $H3$\\

\textbf{S} & \textsc{Scroll} &
Multiple overlapping views on the same file & $H2$ (potential), $H3$ \\

\textbf{DZ} & \textsc{OverlyDeepZoom} &
A sequence of view actions on a specific file or directory, may or may not be followed by patching & $H3$, $H4$ (optional)\\
\hline

\textbf{UR} & \textsc{UnresolvedRetry} & Multiple consecutive failed edits (same file) with no later success & $H1$, $H3$, $H4$\\
\textbf{ER} & \textsc{EditReversion} & Reverting a previous successful edit & $H3$\\
\textbf{NF} & \textsc{StrNotFound} & Edit fails because the specified old string does not exist in the file & $H3$, $H4$ \\
\textbf{NE} & \textsc{NoEffectEdit} & Edit has no effect due to identical old and new strings & $H3$, $H4$\\
\textbf{AT} & \textsc{AmbiguousTarget} & The target string occurs in multiple locations
& $H3$, $H4$ \\

\bottomrule
\end{tabular}
\end{table}

Table~\ref{tab:common-patterns} summarizes the identified anti-patterns, found only in the localization and the patching phase. None of the automatically detected anti-patterns during the validation phase was identified as an inefficiency through manual analysis. Below, we explain these anti-patterns in more detail and with qualitative examples. 


\textbf{RepeatedView} occurs when the agent revisits the same structural region, e.g., a file, multiple times during navigation, reflecting redundant inspections of identical code, often after unsuccessful edits or incomplete localization. As shown in Figure~\ref{fig:graphectory-example}-a, after two failed edit attempts (step~9), the agent reopens \texttt{client.py} (the back edge $10$ to the same \texttt{\small{view}}) for better localization. 

\textbf{ZoomOut} refers to cases where the temporal order of \textit{view} actions contradicts the structural hierarchy, meaning the agent moves from a deeper to a shallower level in the project structure. This pattern indicates a backward navigation,
i.e., the agent realizes it is exploring a wrong subdirectory or file. As shown in Figure~\ref{fig:graphectory-example}-b, after entering the wrong directory \texttt{\small postgres} (step~3), the agent goes back to the parent folder \texttt{\small /testbed/db} (step~4) and then navigates into the correct sibling \texttt{\small /testbed/db/postgresql} (step~5) to locate the buggy file \texttt{\small client.py} (step~6).

\begin{figure}
    \centering
    \includegraphics[width=0.93\linewidth]{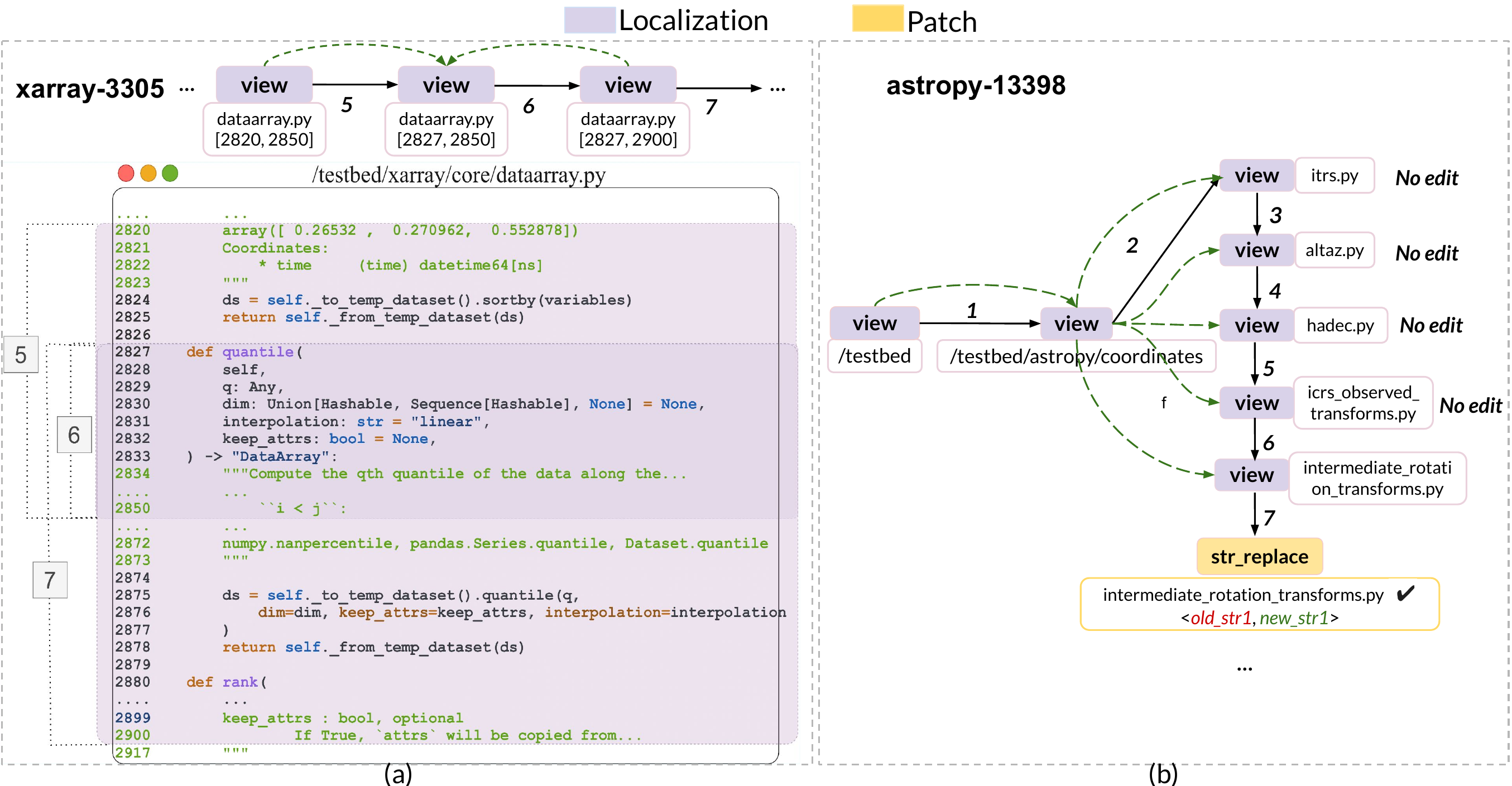}
    \vspace{-8pt}
    \caption{Example of anti-patterns Scroll (a) and OverlyDeepZoom (b)}
    \vspace{-10pt}
    \label{fig:localization_patterns}
\end{figure}

\textbf{Scroll} occurs when multiple \textit{view} actions target overlapping line ranges within the same file. Figure ~\ref{fig:localization_patterns}-a shows \SA\textsubscript{DSK-V3} performs several consecutive \texttt{\small view(path,range)} actions on \texttt{\small dataarray.py}, first inspecting lines $2820-2850$, 
then lines $2827-2850$, 
and finally lines $2827-2900$, 
each overlapping the previous view. Such redundant scrolling occurs when the agent is unaware of function definition boundaries, requiring repeated views to capture the complete function body.

\textbf{OverlyDeepZoom} happens when no \texttt{\small edit} follow \texttt{\small view} \emph{on the same file path}. This can be due to inefficient issue-based localization, causing the agent to explore several files in the same directory until it finds the bug location. As a result, the agent may never find the correct location to edit, make an unsuccessful edit, or successfully edit after an extended time. 
Figure~\ref{fig:localization_patterns}-b shows an instance of last case, where the \SA\textsubscript{DSK-V3}, after narrowing down to the directory \texttt{\small coordinates} (step~2), sequentially inspects four sibling files (\texttt{\small itrs.py}, \texttt{\small altaz.py}, \texttt{\small hadec.py}, and \texttt{\small icrs\_observed\_transforms.py}) to finally finds the buggy location (steps~3–6) and repair (step~7).

\begin{figure}[t]
  \centering
  \includegraphics[width=0.97\linewidth]{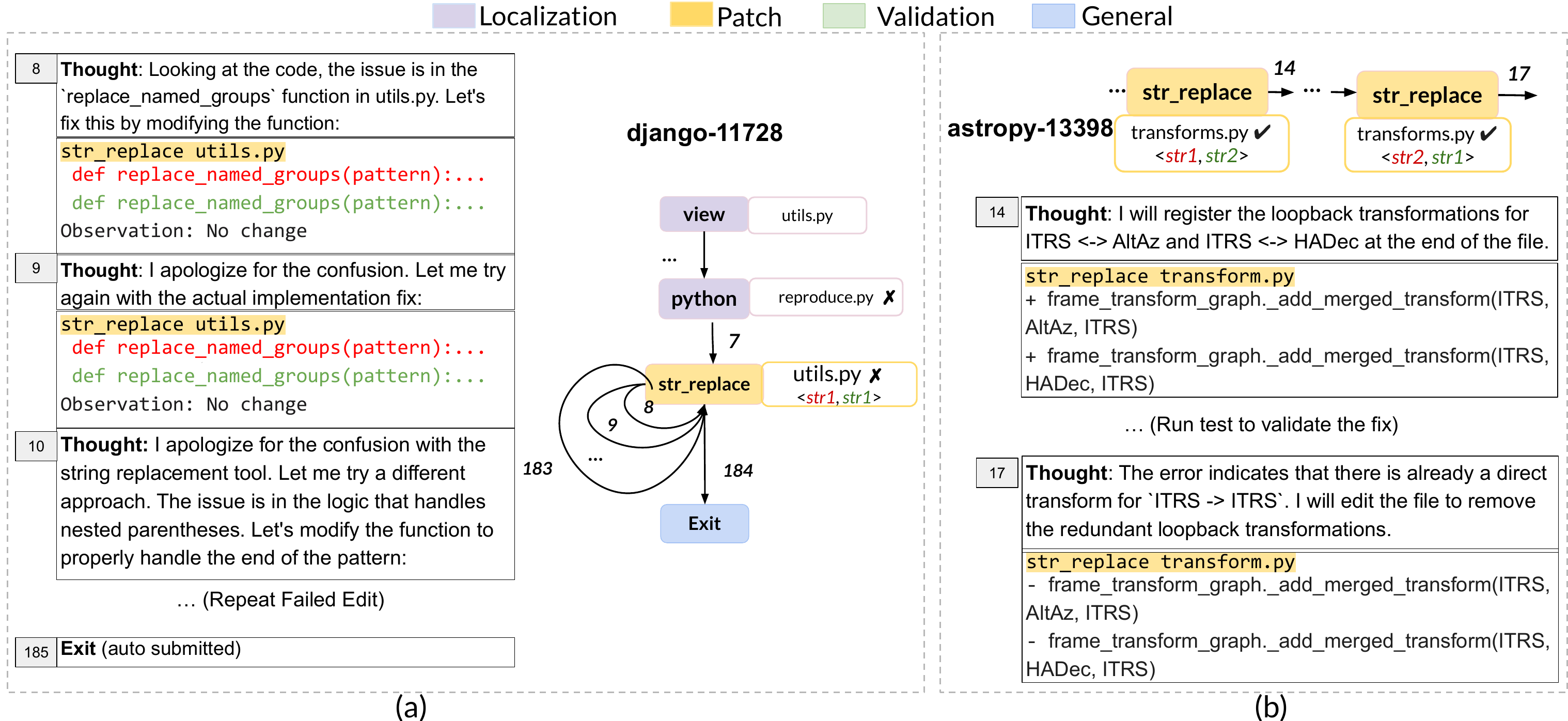}
   \vspace{-5pt}
  \caption{Example of anti-patterns UnresolvedRetry (a) and EditReversion (b)}
  \label{fig:patching_patterns}
  \vspace{-5pt}
\end{figure}

\textbf{UnresolvedRetry} represents the scenario where an agent performs multiple consecutive edits on the same file, but all of which fail to yield a successful modification. Figure~\ref{fig:patching_patterns}-a shows an example of such a case, where \SA\textsubscript{Dev} repeatedly executes \texttt{\small str\_replace} commands on \texttt{\small utils.py} for $183$ times, reaching the cost limit without any success.

\textbf{EditReversion} captures cases where a previously successful edit is later reverted. Figure~\ref{fig:patching_patterns}-b shows \SA\textsubscript{DSK-V3} first modifies \texttt{\small transform.py} (step $14$), leading to a series of test execution failures (steps $15$--$16$, not shown in the figure) and reverting to the original file (step $17$). 

\begin{figure}[t]
  \centering
  \includegraphics[width=0.9\linewidth]{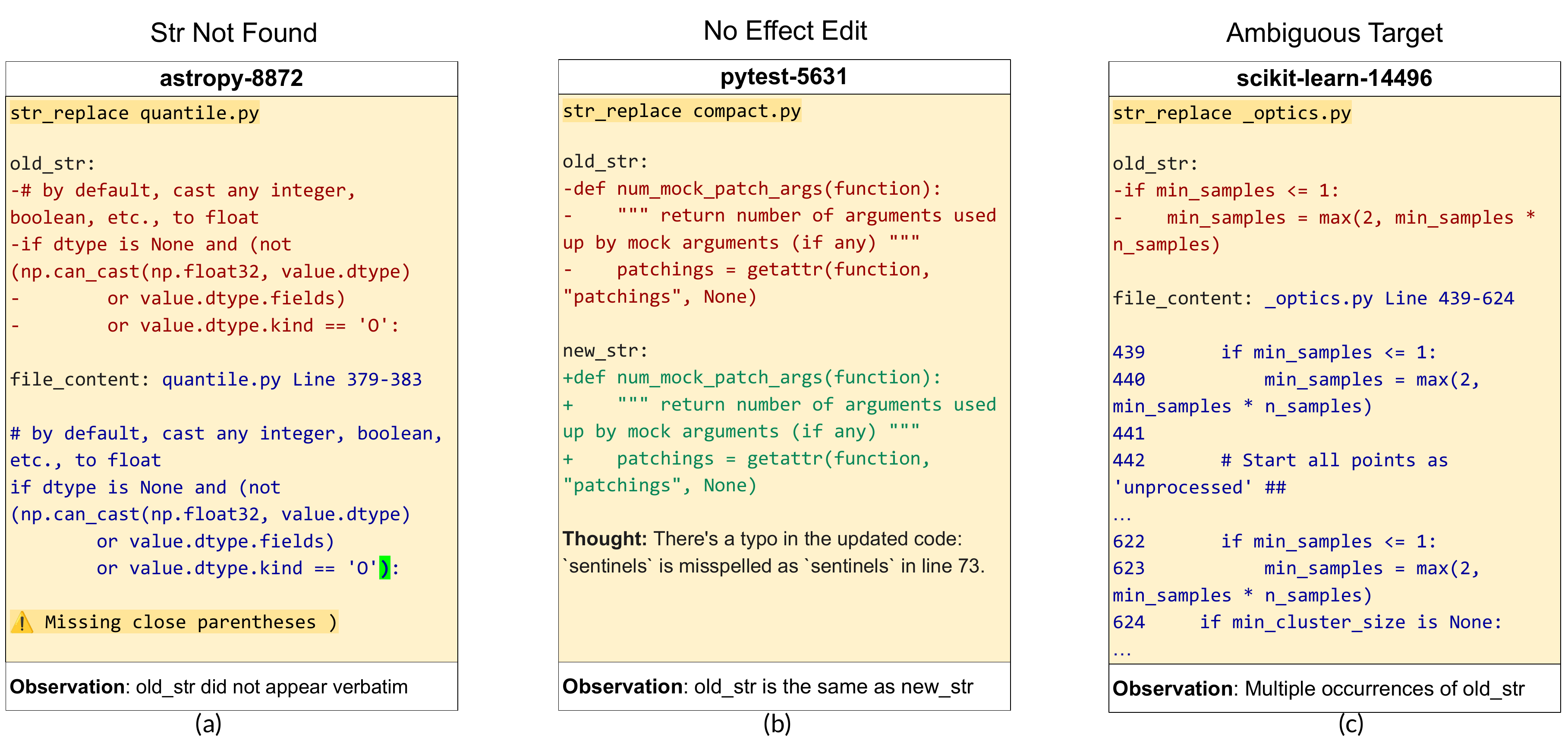}
   \vspace{-5pt}
  \caption{Editor failure modes of (a) \textsc{StrNotFound}, (b) NoEffectEdit, and (c) AmbiguousTarget}
  \vspace{-5pt}
  \label{fig:edit_failure_modes}
\end{figure}

\textbf{StrNotFound} occurs when an edit fails because the specified old string cannot be found in the file (Figure~\ref{fig:edit_failure_modes}-a). Such failures typically happen due to formatting issues or minor text differences. 

\textbf{NoEffectEdit} happens because the old and new strings are identical, producing a meaningless replacement. Figure~\ref{fig:edit_failure_modes}-b illustrates an example of this, where the command replaces a string with itself, resulting in no change to the file and wasting execution cost. 

\textbf{AmbiguousTarget} captures cases where the specified old string appears multiple times in the file, leading to ambiguity about which occurrence should be replaced. Figure~\ref{fig:edit_failure_modes}-c illustrates an example of this, where multiple identical \texttt{\small if} statements cause the editor to fail in applying the replacement deterministically, showing a key limitation of string-based editing mechanisms.

\begin{figure}[t]
  \centering
  \includegraphics[width=0.88\linewidth]{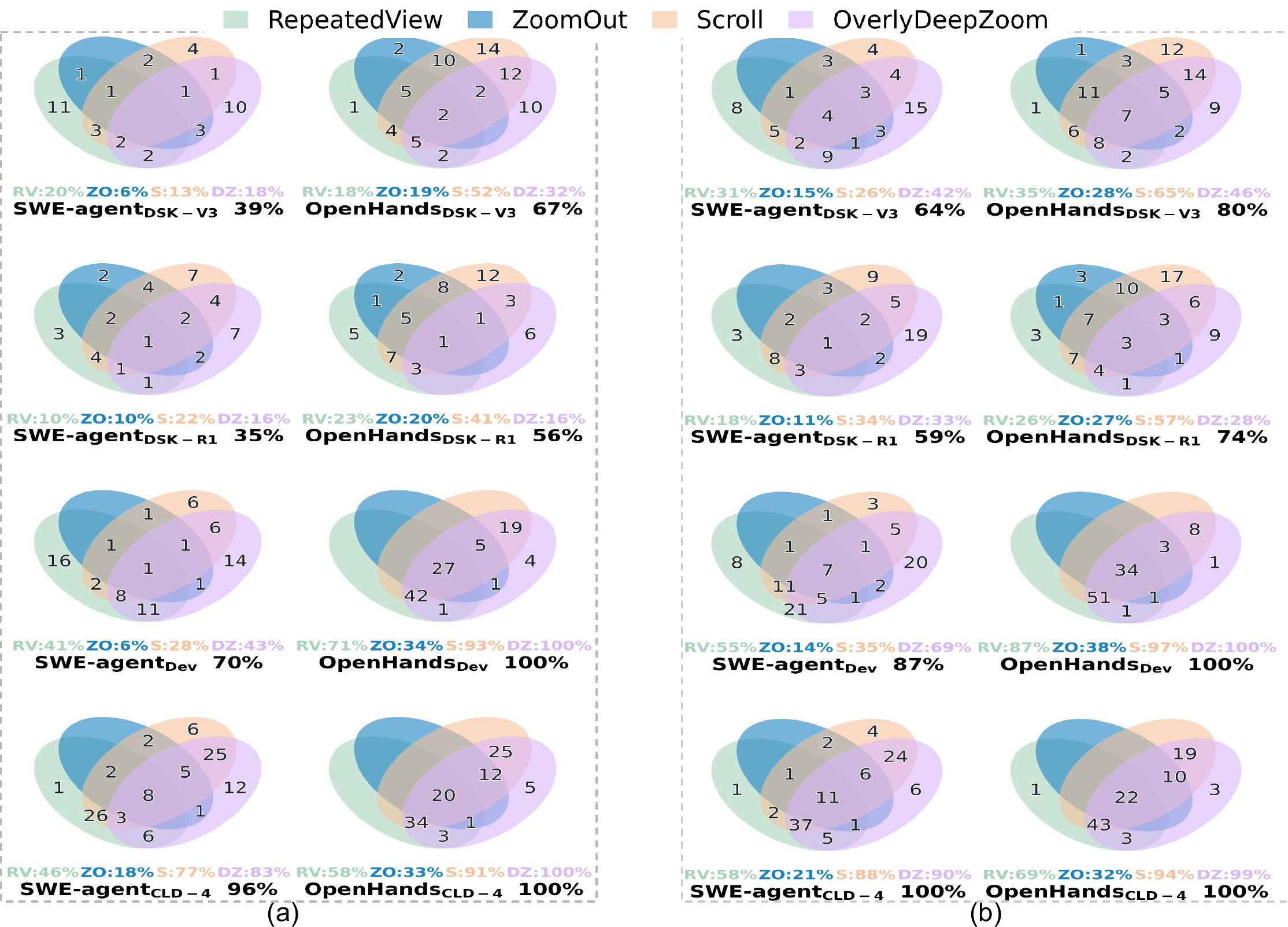}
  \vspace{-8pt}
  \caption{Localization inefficiency patterns among Resolved (a) and Unresolved (b) instances}
  \vspace{-5pt}
  \label{fig:loc-ineff}
\end{figure}

We formalized the identified patterns and searched them in $4000$ studied \approach. Figure~\ref {fig:loc-ineff} summarizes the distribution of four localization inefficiency patterns across all agents and models for resolved (Figure~\ref {fig:loc-ineff}-a) and unresolved instances (Figure~\ref {fig:loc-ineff}-b). The percentages below the Venn diagrams show the instances that had at least one of the localization inefficiency patterns.

An important observation is that inefficiencies, although to a lesser extent, are prevalent in resolved instances as well. This highlights the importance of process-centric analysis to raise awareness about the behavior of agents. Unresolved cases often involve multiple inefficiency patterns within a single trajectory. Interestingly, stronger models such as Claude Sonnet 4 achieve higher success rates yet exhibit more of these patterns. Manual analysis suggests that the model tends to engage in extended internal reasoning before producing a solution. Although this can be helpful when solving difficult problems (as evidenced by the model's higher success rate), it also leads to inefficiency when the additional steps do not contribute to better solutions.

\begin{figure}[t]
  \centering
  \includegraphics[width=0.9\linewidth]{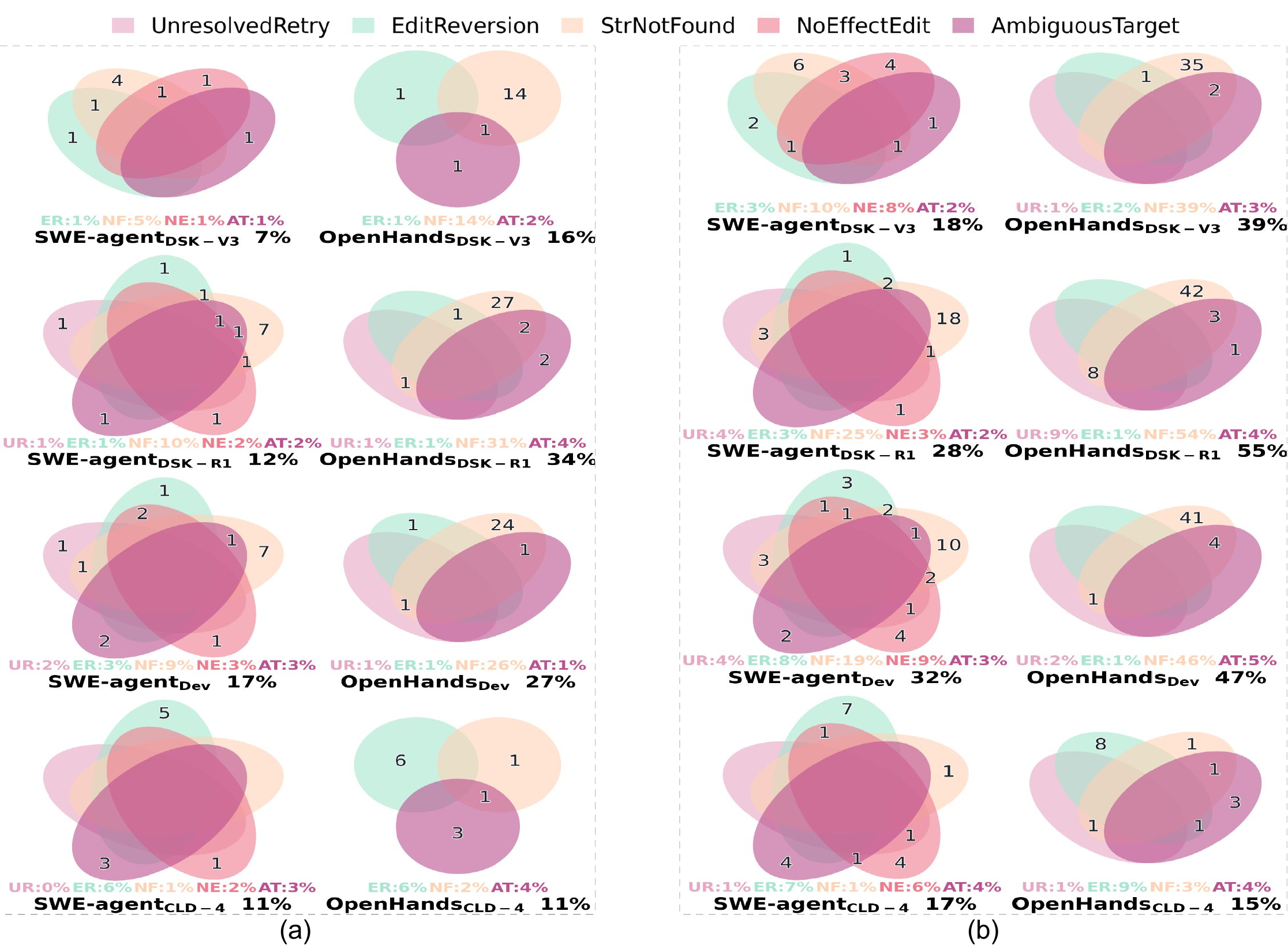}
   \vspace{-5pt}
  \caption{Patching inefficiency patterns among Resolved (a) and Unresolved (b) instances}
   \vspace{-5pt}
  \label{fig:edit_venn}
\end{figure}

Figure~\ref{fig:edit_venn} shows the distribution and overlap of patching inefficiencies for resolved (Figure~\ref{fig:edit_venn}-a) and unresolved (Figure~\ref{fig:edit_venn}-b) instances. In resolved instances, anti-patterns appear less frequently across categories than in unresolved instances.
\emph{Unresolved instances have more inefficient patterns}, particularly \emph{StrNotFound} and \emph{EditReversion}, indicating the agents often fail to locate target code or oscillate between inconsistent edits. \OH variants tend to show stronger accumulation of inefficiencies than \SA, and \emph{weaker models suffer from more editing failures} than stronger models like Claude-4. Overall, existing agents still struggle with simplistic string-based editing. Future repair tools that incorporate syntax-aware modifications and stronger planning mechanisms may strike a better balance between accuracy and efficiency in resolving issues.


\begin{findingsbox}
\textbf{Findings.} Inefficiencies are common even in resolved runs, though to a lesser extent. Unresolved runs often accumulate multiple localization or editing anti-patterns within a single trajectory. Despite their higher success rate, stronger models (e.g., Claude-4) exhibit more inefficiencies, likely due to their greater internal exploration. 
\end{findingsbox}

\subsection{Online Monitoring and Intervention}
\label{sec:mr-monitor}


\textcolor{\MRColor}{To investigate the effectiveness of the online monitoring and intervention, we implement it on top of \SA framework. The current implementation monitors partial \approach and \lang for three classes of problematic behavior during execution: (1) Plan violation (e.g., skipping validation after patching, or patching before localization); (2) Oscillatory behavior (H1 from \S \ref{subsub:online-monitoring-analysis}), detected as cycles or repeated cycles in the \approach (e.g., self-loops or repeated loops without actual progress); and (3) Prolonged stagnation in a single phase (H3 from \S \ref{subsub:online-monitoring-analysis}), signaling that the agent may be stuck (e.g., spending at least $5$ steps in navigation without any edit).} 


\textcolor{\MRColor}{At each step, the monitor evaluates the partial \approach/\lang against these rules. 
For plan violation and oscillation cases, the intervention component suppresses the current step, i.e., drops the thought–action–observation, and returns only a diagnostic message. Accordingly, it updates the \approach and \lang consistent with the trajectory. 
In case of plan violations, the monitor generates a message using the following template: \emph{``Plan violation: \{violation\_type\}. Expected to \{expected\_phase\}. Skipped phases: \{skipped\_phases\}. Follow the intended workflow to ensure thorough problem solving.''} The template is instantiated based on the detected violation. For oscillatory behavior, the monitor returns a guidance message such as: "A repeating cycle of actions is detected. Summarize the failing behavior and choose one decisive next step." For prolonged stagnation, the intervention component keeps the current step\footnote{The rationale is twofold. First, rolling back due to prolonged stagnation is more difficult than for the other rules: in those cases, the culprit is the most recent action, whereas for prolonged stagnation, it is unclear. Second, the history may be useful for determining the next action when accompanied by the diagnostic message.} and appends a targeted corrective instruction as a user message, starting with "You may be stuck in \{phase\}." and customized depending on the applicable phase. For example, if the phase is localization, the message includes "Focus on relevant code and run tests to reproduce and localize the bug if necessary."}

\begin{table}[t]
\centering
\small
\caption{Effect of online monitoring and intervention on reproducible problematic instances. OMI refers to a setting with integrated online monitoring and intervention.}
\vspace{-10pt}
\label{tab:monitor}
\begin{tabular}{llccccc}
\toprule
Model & Setting & \#Instances & Resolution Rate (\%) & Trajectory Steps & Oscillation (\%) & Plan Violation (\%)  \\
\midrule
DSK-R1 & Vanilla & \multirow{3}{*}{17} & 11.8 & 38.1 & 100.0 & 52.9  \\
                 & OMI  &                     & 35.3 & 26.6 & 5.9   & 17.6 \\
                 & $\Delta$ &                   & +23.5 & $-$11.5 & $-$94.1 & $-$35.3\\
\midrule
DSK-V3      & Vanilla & \multirow{3}{*}{23} & 8.7  & 41.4 & 100.0 & 73.9 \\
                 & OMI  &                     & 21.7 & 28.3 & 0.0   & 17.4 \\
                 & $\Delta$ &                   & +13.0 & $-$13.1 & $-$100.0 & $-$56.5 \\
\midrule
Dev   & Vanilla & \multirow{3}{*}{46} & 10.9 & 99.5 & 100.0 & 37.0 \\
                 & OMI  &                     & 17.8 & 39.6 & 8.9   & 8.9  \\
                 & $\Delta$ &                   & +6.9 & $-$59.9 & $-$91.1 & $-$28.1 \\
\bottomrule
\end{tabular}
\end{table}

\textcolor{\MRColor}{ 
We re-executed \SA with integrated online monitoring and intervention (OMI setting) on SWE-Bench Verified instances with trajectory-level issues under the vanilla setting from RQ1. To minimize the impact of non-determinism on the results, we repeated the vanilla SWE-agent runs three times on previously problematic instances and retained only those in which the issues persisted in at least two runs. This yielded $86$ reproducible problematic trajectories (out of $153$). We compare the OMI and vanilla settings (Table~\ref{tab:monitor}) concerning the percentage of resolved instances, number of trajectory steps, percentage trajectories with oscillation, and percentage of trajectories with plan violation. The last two specifically highlight the effectiveness of the intervention, i.e., the extent to which trajectory rollback and the diagnostic message helped the agent avoid the same mistake. 
Under the vanilla setting, an instance is considered resolved only if all reproducible runs resolve it. The number of steps is also the average across all reproducible runs.}

\textcolor{\MRColor}{Table~\ref{tab:monitor} reports the results of this experiment.
Overall, online monitoring and intervention substantially reduce oscillatory behavior across all models (by over $90\%$), and consistently shorten trajectories. At the same time, it leads to higher resolution rates, with improvements ranging from $6.9\%$ to $23.5\%$ on studied instances. This proves the effectiveness of online process-centric analysis, resulting in efficient trajectories with a higher success rate at a lower cost. 
Plan compliance of agents is improved as well, although to a lesser extent compared to oscillatory behavior. We speculate this is because breaking oscillatory loops provides a clear corrective signal that agents readily follow. In contrast, plan compliance requires agents to override their judgment about the problem-solving stage. Even when instructed to follow the intended workflow, agents may act based on their belief that the task is complete. For example, they might still submit a patch without validation, despite being told to test, if they believe the fix is sufficient and additional testing is unnecessary.}

\vspace{-3pt}
\begin{findingsbox}
    \textbf{Findings.} Online process-centric analysis and intervention enable timely detection and repair of trajectory issues, significantly shortening trajectories and improving resolution rates.
\end{findingsbox}

\section{Related Work}
\label{section:related-work}


In the last few years, numerous autonomous agents have emerged for automated issue resolution and software engineering tasks~\citep{SWE-agent, OpenHands, aider, chen2024coder, moatless, liu2024marscode, composio_agent}.  
\approach is designed to be generic and can be used to represent and analyze different agentic programming systems.
Recent works have called for richer evaluation paradigms that account for efficiency, robustness, and reasoning quality~\cite{wang2024survey,park2023generative,shinn2023reflexion,liu2023agentbench,liu2025empirical,ceka2025understanding}. 
\textcolor{\MRColor}{\citep{bouzenia2025understanding} analyzes thought-action-result trajectories of software agents to study their interaction patterns and recurring behaviors.}
Others~\citep{deshpande2025trail, cemri2025multi} have explored the failure modes of SWE-agents by analyzing their trajectories.
\citet{deshpande2025trail} release TRAIL (Trace Reasoning and Agentic Issue Localization), a taxonomy of errors and a corresponding benchmark of 148 large human-annotated traces
from GAIA and SWE-Bench Lite. \citet{cemri2025multi} propose MAST (Multi-Agent System Failure Taxonomy)
for multi-agent systems, that, despite our technique, require human annotations for analysis.

\citet{ceka2025understanding} builds a taxonomy of decision-making
pathways based on agents' execution traces, categorized by project understanding, context understanding, and patching and testing actions. Their bug-localization analysis was performed manually on 20 issues from 500 in SWE-Bench Verified. \citet{liu2025empirical} develops a taxonomy of LLM failure modes for SWE tasks comprising 3 primary phases, 9 main categories, and 25 fine-grained subcategories. They identify tool usage failures and pinpoint issues such as unproductive iterative loops in agent-based tools. However, their underlying cause analysis is conducted manually on 150/500 issues.

A common theme across prior work is their reliance on manual annotations, thereby limiting the number of annotations available for further analysis. \approach does not rely on manual annotations for the entire process, and once the phase map has been updated to incorporate any new tools, we can proceed to \emph{phase labeling} and analyze any number of trajectories (thousands) across any number of agents using \lang. We will support additional agents upon request as the industry evolves from single-agent to multi-agent systems, e.g, multi-agent research system in Claude.ai \citep{claudemultiagentresearchsystem}
\vspace{-10pt}
\section{Conclusion}

This paper presents \approach, a structured representation of agentic trajectories to move beyond outcome-centric evaluation. Our experiments of analyzing two agentic programming frameworks demonstrate that \approach enables richer forms of analysis and paves the way for novel evaluation metrics that reflect not only the agent's success, but how it proceeds through a task. This perspective highlights opportunities for designing more efficient and robust agentic systems. We believe \approach opens several research directions, which we plan to explore as the next step: Developing structure-aware program analysis tools for symbolic navigation, targeted context retrieval, and AST-based editing to reduce the inefficiencies exposed by \approach. 

\vspace{-5pt}
\section{Acknowledgments}

This work is supported by the IBM-Illinois Discovery Accelerator Institute and NSF CCF-2238045 grants. We thank Dr. Martin Hirzel for his valuable feedback. We also thank the anonymous reviewers for their comments, which helped make this work stronger.

\vspace{-5pt}
\section{Data-Availability Statement} 
\label{sec:data-availability}

Our artifacts are publicly available~\cite{website}, including (1) the implementation of \approach and \lang, which we hope the researchers and practitioners use for better evaluation of agentic programming systems, (2) the implementation and all analyses reported in the paper, (3) a dataset of \approach and their corresponding raw trajectories from all $4000$ runs ($500$ SWE-bench Verified instances $\times$ 8 $\langle$agent, model$\rangle$ pairs), \textcolor{\MRColor}{and (4) trajectories generated through online monitoring and intervention.} 
\balance
\bibliographystyle{ACM-Reference-Format}
\bibliography{references}

\end{document}